%% file: main_final.tex
\documentclass{dsj}

\usepackage{natbib}
\usepackage{amsfonts,amsmath,amsthm}
\newtheorem{property}{Property}
\usepackage{makecell} %
\usepackage{booktabs,array,multirow}
\usepackage{siunitx}

\begin{document}

\footer{Footer Text}
\submitted{27 January 2026}

\title{Functional requirements decomposition in set-based design}

\author[1]{Minghui Sun}
\author[2]{Zhaoyang Chen}
\author[3]{Georgios Bakirtzis}
\author[4]{Hassan Jafarzadeh}
\author[2]{Cody Fleming\email{flemingc@iastate.edu}}

\address[1]{School of Automation, Nanjing University of Information Science and Technology}
\address[2]{Iowa State University}
\address[3]{LTCI, Télécom Paris, Institut Polytechnique de Paris}
\address[4]{University of Virginia}

\maketitle

\input{body_final_26Jan}
\bibliographystyle{plainnat}

\bibliography{references.bib}

\end{document}

%% file: body_final_26Jan.tex
\begin{abstract}
Designing systems is typically uncertain and ambiguous at early stages. Set-based design supports alternative exploration and gradual uncertainty reduction during the early lifecycle, making it practical for complex systems design. In parallel, the functional requirements decomposition helps to advance the design incrementally. However, current literature on set-based design lacks formal guidance in how to decompose functional requirements. To bridge this gap, we introduce a four-step method to decompose functional requirements for set-based design hierarchically. We systematically define, reason, and narrow the sets,  breaking down the functional requirements into formal sub-requirements. This method allows parallel abstraction, ensuring the resulting system satisfies the top-level functional requirements.

\end{abstract}

\section{Introduction}\label{sec:intro}

Modern engineered systems are usually developed in an interdisciplinary approach focused on defining customer (or stakeholder needs), necessary functionality, and documenting requirements early in the development cycle. These requirements are then used to proceed with design synthesis and system validation while considering myriad factors such as operations, cost, schedule, performance, human support, and so on \citep{booton2008development,eisner2011essentials,kapurch2010nasa}. Design complexity and the time and cost constraints emanating from competitive market needs often require critical decisions to be undertaken very early in the design process \citep{al2024design}. A critical aspect of this approach is the identification and decomposition of technical requirements which ultimately form the foundation for the architecture, design, integration, and verification of the resulting system \citep{incose2023incose}. 

One of the challenges in requirements decomposition is the inherent uncertainty and ambiguity in the early stages of designing systems \citep{schrader1993choice}. For example, design parameters can exhibit high levels of uncertainty, and such uncertainty—both aleatoric and epistemic—must be appropriately characterized to enable its impact to be assessed and subsequently mitigated or tolerated; requirements and their decomposition methods—often expressed in natural language—are notoriously ambiguous, leading to confusion and errors during the decomposition process; the deliberate ambiguity embedded in high-level requirements, intended to allow flexibility for lower-level creative implementations, introduces additional challenges in maintaining transparency and traceability during cross-level ambiguity refinement.
Moreover, developing complex systems (e.g., commercial aircraft) usually involves a number of (globally) distributed design teams. The fact that independent design teams (for subsystems and components) operate as semi-autonomous units with loose inter-team coordination, particularly in large projects  \citep{rismiller2023using} makes addressing this challenge even more daunting. 
It has been noted that system failures are equivalent to requirements failures \citep{leveson1993investigation,leveson2016engineering}, with system accidents or other failures often due to missing, flawed, or inconsistent requirements \citep{fleming2015including,fleming2015integrating}. 

Set-based design (SBD) \citep{ward1989theory,Toche_Pellerin_Fortin_2020} arises as a potential candidate for the above challenges and is particularly suited in several ways as it supports robust development of design alternatives, uncertainty reduction and resolution \citep{shallcross2021quantitative}, and subsystem autonomy and optimization, i.e. the ability for subsystems and parts to be developed separately and in parallel \citep{ghosh2014set}. Function decomposition features such set-based reasoning \citep{ghosh2014set}. In SBD, the overall system design problem is typically divided into multiple distinct disciplines, each exploring their own sets of possibilities \citep{specking2018literature}. Teams of engineers develop a set of design alternatives in parallel at different levels of abstraction and progressively narrow down the prospective alternatives based on additional information until a final solution is reached \citep{sobek1999toyota}. SBD is defined in greater detail in Section \ref{sec:background}. Although the aforementioned requirements engineering challenges broadly, and function decomposition process specifically, may not be nominally related to SBD, function decomposition has been noted to feature set-based reasoning \citep{ghosh2014set}.

To tackle uncertainty and ambiguity, one of the most effective approaches is to formalize. However, there has been limited formal guidance in the SBD literature on how to define, reason, and narrow sets while improving the level of abstraction of the design \citep{dullen2021survey,specking2018literature}. Therefore, we propose a four-step method in this paper that applies set-based reasoning to, systematically and formally define, reason, and narrow the sets, eventually decomposing the functional requirements into sub-requirements. Specifically, this paper has two main objectives:
\begin{enumerate}
    \item Creating a formal method to decompose functional requirements based on SBD principles to tackle ambiguity and uncertainty in requirements decomposition.
    \item Ensuring Objective 1 supports parallel development, i.e., autonomy among lower-level design teams.
\end{enumerate}

For Objective 1, in addition to addressing a fundamental challenge in requirements generation, our method makes two significant contributions to the SBD literature:

\begin{enumerate}
    \item Most SBD approaches formulate functional requirements as ranges within performance spaces, which is applicable to many mechanical components. However, for higher-level systems that may involve a collection of mechanical systems as well as software and other types of components, functions are typically defined as generic transformations between inputs and outputs. Accordingly, functional requirements must be defined as mappings between input and output ranges. Our method focuses on this latter formulation, addressing requirements development for complex systems and complementing existing SBD literature.
    \item In the design community there are two ways to address uncertainties  \citep{eckert2019design}: ``buffer'' (the portion of parameter values that compensates for uncertainties) and ``excess'' (the value over and above any allowances for uncertainties). The current SBD literature does not explicitly distinguish between buffer and excess when addressing uncertainties, lacking specificity in robustness claims. Our method clearly distinguishes between buffer and excess in their respective steps, where buffer is used to reduce controllable uncertainties, and excess accommodates the possibility of underestimated initial uncertainties.
\end{enumerate}

For Objective 2: SBD claims that the autonomy of design teams is an advantage \citep{matthews2014bayesian}, but very little SBD work provides a priori proof of this property. In fact, one cannot rigorously justify such a claim without an explicit underlying formalism. Based on the formalism defined in this paper, we can prove that when functional requirements are decomposed in a specific way (i.e., composable and refinement), individual design teams can work independently and the resulting system will satisfy the top-level functional requirements, realizing the promise of parallel development. 

\begin{figure}[!t]
\centerline{
\includegraphics[width=\textwidth]{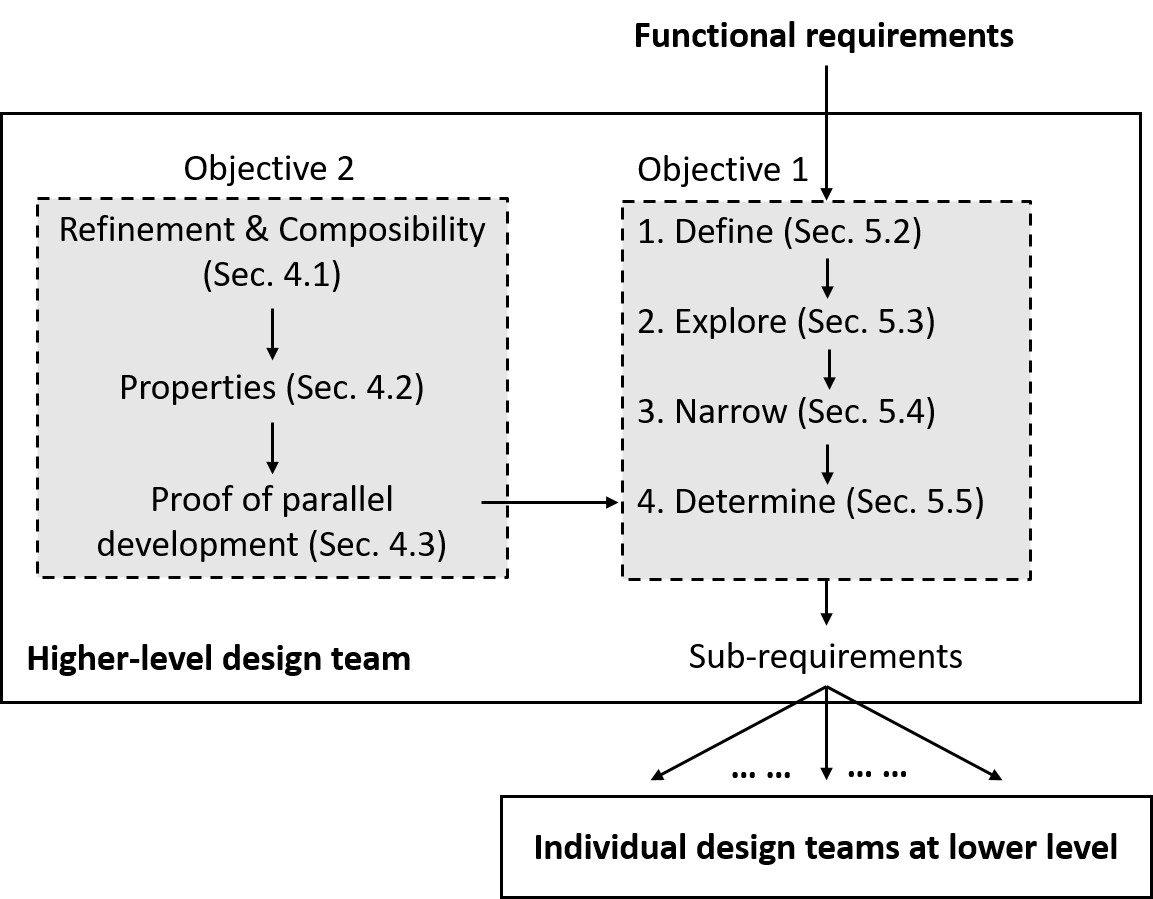}
}
\caption{The overall structure of the proposed approach.}
\label{fig:structure}
\end{figure}

Fig.\ref{fig:structure} is an overall description of the structure of the proposed approach. After introducing the background information and the preliminaries (Section \ref{sec:background} and \ref{sec:pre}), we address Objective 2 first in Section \ref{sec:pd} (the left half of Fig.\ref{fig:structure}). It was concluded that to ensure parallel development, the sub-requirements at each level of
abstraction must be \emph{composable} and able to \emph{refine} the higher-level functional requirements as a whole. This finding will be treated as the constraints of the optimization problem of the requirements decomposition process. After that, we explain the requirements decomposition process based on SBD principles (the right half of Fig.\ref{fig:structure}) in Section \ref{sec:decompose}. Given the functional requirements from the customers or a higher-level design team, a four-step process is applied to define, explore and narrow the possible space of alternatives (Section \ref{sec:architecture}--\ref{sec:dsexp}) and eventually determine the sub-requirements that will be assigned to design teams at the lower levels through an optimization program (Section \ref{sec:reqdev}). We demonstrate the method with an example of designing a cruise control system based on existing computational tools in Section \ref{sec:case}. 

\section{Background}\label{sec:background}
\subsection{Set-based Design (SBD)}
In product development, two main strategies exist for selecting a concept: Point-Based Design (PBD) and Set-Based Design (SBD). PBD involves choosing a single “best” option early from all possibilities, whereas SBD progressively eliminates infeasible or less attractive alternatives until only one solution remains \citep{morgan2020toyota}. PBD often causes problems because decisions are made too early, based on assumptions or incomplete knowledge, leading to rework, correction loops, and risk aversion that can stifle innovation \citep{majerus2017lean}.

SBD addresses these issues by maintaining multiple feasible options longer and narrowing them down as knowledge grows. SBD is a structured engineering methodology that emphasizes exploring a wide spectrum of design alternatives rather than fixating on a single solution from the outset. Originating in lean product development, most notably at Toyota \citep{ward1995second}, this approach promotes delaying commitment until sufficient knowledge is amassed, enabling more robust trade-offs and reducing costly rework \citep{verma2023systems}. In SBD, cross-disciplinary teams concurrently maintain multiple feasible design sets, gradually converging through feasibility assessment and intersection of acceptable solution spaces \citep{wikisebok}. Unlike traditional point-based or sequential design methods, SBD fosters greater parallelism, improved organizational learning, and the capacity to manage uncertainty—especially in early stages with loosely defined requirements or conflicting constraints. Three principles guide SBD: map the design space, integrate by intersection, and establish viability before committing \citep{sobek1999toyota}. Mapping the design space involves defining feasibility regions, exploring trade-offs via multiple alternatives, and communicating sets \citep{temam2013set}. Integration by intersection ensures that acceptable solutions lie in the overlapping feasible sets of all subsystems, promoting compatibility. Finally, narrowing occurs gradually by refining alternatives, remaining within feasible regions, and managing uncertainty at decision gates. Together, these principles make SBD a structured approach for fostering robustness and innovation while avoiding premature commitment.
A line of research was conducted recently to examine by applying computational methods the benefits and drawbacks of SBD, specifically how SBD interacts with various team and problem structures, using problem coupling and varying timespan to further elucidate the impacts of the process \citep{rismiller2024understanding,rismiller2023exploring}. The results suggest that SBD lowers the need for communication and rework on coupled problems, allowing projects to achieve greater concurrency and break through barriers in the design space to create new designs \citep{rismiller2023using}. 

Recent studies have extended SBD to complex engineering systems, integrating model-based systems engineering (MBSE) to systematically evaluate design options using computational models \citep{specking2018early}. Advanced applications in automotive, aerospace and construction show that combining SBD with optimization and tradespace exploration enables designers to handle uncertainty, balance conflicting requirements, and improve performance early in the development process \citep{Oyama2024,van2024set,Ishikawa2024,TSUDA2024}. Recent reviews \citep{oliveira2024review,castaneda2023set} further confirm that, while theoretical development is still evolving, growing research underscores SBD’s growing relevance in enabling lean, flexible, and resilient design processes.

\subsection{SBD for Requirements}

\subsubsection{Qualitative SBD Approaches}\label{sec:qualitative}
It was pointed out that ``there is limited SBD research contributing to requirements development \citep{shallcross2020set}'' and SBD methodologies applied to complex systems are mostly qualitative \citep{shallcross2021quantitative}.
Enhanced Function-Means Modeling (EF-M) is one such method used for function modeling \citep{muller2019enhanced}. It has been combined with SBD to manage platform-based product family design \citep{raudberget2014combining}. Functional decomposition and solutions are generated according to EF-M, with mappings from Functional Requirements (FR) to Design Solutions (DS) at each level of abstraction \citep{levandowski2014set}. These mappings align conceptually with our proposed method. A systematic, knowledge-focused framework was proposed to guide early-phase design exploration in a set-based, front-loaded manner \citep{gamage2024set}. This framework offers a step-by-step approach to identifying knowledge requirements in design spaces characterized by multiple objectives and design parameters, ensuring better-informed decisions and fostering innovation in complex product development.

A ``wayfaring'' model for set-based requirement generation was introduced to discover critical functionalities and create dynamic requirements \citep{kriesi2016creating}. This model uses prototyping to guide the design process from initial concept to final product, minimizing tooling and production costs. Model-Based Requirements and Architecture (MBRMA), a novel set-based approach, was developed to filter out weak or costly solutions and assist system engineers in trade-off analysis \citep{borchani2019integrating}. Although this approach defines a mathematical formalism, it lacks specific guidance on requirement decomposition.
A framework based on set-based engineering allows for building reusable and adaptable engineering methods \citep{vallhagen2013framework}. These ``virtual methods'' help create and validate early phase design requirements, ensuring the introduction of novel technologies without compromising risk and cost. Other frameworks include CONGA \citep{al2013capturing}, DMIV \citep{ammar2017architectural}, MBSS \citep{yvars2022towards}, RR-LeanPD model \citep{al2013transformation}, and a combination of V-model and SBCE \citep{al2009set}.

One limitation of these qualitative approaches is the ambiguity surrounding the concrete activities needed for requirements development. This ambiguity makes it challenging for industry practitioners to replicate these methods. A formal method  provides precise, repeatable instructions for decomposing functional requirements and assigning them to lower abstraction levels.

\subsubsection{Quantitative SBD Approaches}

Quantitative techniques for design space exploration in SBD often feature hierarchical decomposition. For example, SBD was applied to the design of a downhole module to validate a laboratory-developed method in an actual industry setting \citep{madhavan2008industrial}. The module was decomposed into chassis and bumper subsystems, with design targets assigned based on a downhole assembly impact model. A set-based approach was used for a multilevel design problem of negative stiffness metamaterials, progressing from macro-level to micro-level \citep{matthews2016hierarchical}. Both studies feature hierarchical decomposition but differ from our problem formulation by treating higher-level design spaces as requirements for lower-level design.

Jansson et al. combined SBD with axiomatic design to evaluate multiple design alternatives against functional requirements \citep{jansson2013requirements}. This work mapped higher-level design spaces to lower-level functional requirements but did not detail the derivation process. Shahan and Seepersad proposed a set-based approach to collaborative design, using Bayesian networks to identify compatibilities and conflicts among distributed subsystems \citep{shahan2009bayesian}. However, this approach focused on optimizing top-level objectives rather than decomposing requirements.

Airplane design requirements can be decomposed into design parameters for various components using SBD principles \citep{kerga2014serious}. Although educational, this game effectively demonstrated the derivation of lower-level design parameter ranges from higher-level parameters.  Panchal et al. presented an interval-based constraint satisfaction method for decentralized, collaborative multifunctional design, systematically reducing interval-based design variables to satisfy requirements \citep{panchal2007interval}. SBD was also used to inform system requirements and evaluate design options for a UAV, demonstrating comprehensive tradespace exploration and valuable insights into requirement development \citep{parnell2019using, specking2021quantitative}.

In summary, these approaches treat functional requirements as ranges of performance space variables, unlike our method, which defines requirements as mappings between design and performance spaces. This distinction significantly impacts the requirements decomposition process, with our method focusing on mapping from higher-level function mappings to sub-function mappings.

\subsection{Design Uncertainty}
\subsubsection{The SBD Approach}
SBD is known for its robustness to design uncertainty \citep{costa2003iteration,terwiesch2002exchanging,bogus2006strategies}. The U.S. Naval Sea Systems Command highlighted SBD's suitability for design problems with conflicting requirements and high uncertainty \citep{parker2017set}. Trueworthy corroborated this, noting SBD's flexibility in adjusting to changing requirements \citep{trueworthy2019set}. Specifically, flexibility is implemented
by initially setting up broad specifications and loose constraints, which can be narrowed down as details emerge \citep{al2024design}. 

An approach incorporating Bayesian network classifiers defined ``design flexibility'' as the subspace size producing satisfactory designs and ``performance flexibility'' as the feasible performance space size relative to the desired space \citep{matthews2014bayesian}.
An SBD methodology was proposed to obtain scalable optimal solutions that satisfy changing requirements through remanufacturing, demonstrated on a structural aeroengine component \citep{al2021scalable}. 
Another line of work combines SBD with platform-based design to preserve design bandwidth, allowing flexibility in different products \citep{muller2020lessons}. A dynamic platform modeling approach represented production variety streams, reducing the risk of late and costly modifications \citep{landahl2021dynamic}.  Modeling platform concepts in early phases and eliminating undesired regions of the design space were described by \citep{levandowski2014set, levandowski2016modularization, levandowski2013using}.
In addition,  McKenney et al. conducted a design experiment showing that delaying decisions using SBD increases adaptability to requirement changes \citep{mckenney2011adapting}. A method called dynamically constrained SBD provided a dynamic map for feasible design space under varying requirements using parametric constraint sensitivity analysis and convex hull techniques \citep{kizer2014set}. This method identified a robust feasible design space and a flexible family of solutions. A hybrid agent approach for set-based conceptual ship design found the process robust to intermediate design errors \citep{parsons1999hybrid}. Similar work can be found in \citep{ross2017determining} and \citep{rapp2018product}.

According to Tackett et al., ``excess'' is ``the quantity of surplus in a system once the necessities are met \citep{tackett2014model}.'' Eckert et al. later defined ``excess'' as ``the value over and above any allowances for uncertainties'' and ``buffer'' as ``the portion of parameter values that compensates for uncertainties \citep{eckert2019design} .'' Current SBD literature lacks an explicit distinction between buffer and excess when addressing uncertainties, reducing specificity in robustness claims.

\subsubsection{The Margin-based Approach}
The aforementioned concepts of ``buffer'' and ``excess'' stem from a line of research called ``design margin'' \citep{eckert2019design}.  
In the seminal papers \citep{clarkson2004predicting,eckert2004change}, margin was a concept closely related to change and tolerance of changes. 
In more recent years, we have seen significant movements in this field. El Fassi et al. proposed a graph-theoretical approach--an assumption network--to support the allocation and management of design margins \citep{el2020assumption}. A Design Belief Network (DBN) was built to help designers determine whether the changes can be absorbed without cascading modifications downstream.
A method called margin value method (MVM) to analyze an engineering design, localize the 
excess margin, and quantify it considering change absorption potential in relation to design performance deterioration \citep{brahma2020margin}. MVM was then 
adapted in the context of uncertainty in input 
specifications so that undesirable effects of 
margins can be minimized while preventing change propagation \citep{brahma2022use}. Later, the same team expanded this method to evaluate sets
of concepts that are combinatorially generated from an enhanced function-means tree by using surrogate models and novel
metrics for evaluating different conceptual alternatives \citep{al2024design}. 

``Resilient objects'' provide passive protection against disruptive events so that the
necessity for complex margins can be reduced at system interfaces \citep{panarotto2024local}. It offered a new perspective
on how to solve the trade-offs between resilience and complexity when addressing uncertainties in the dynamic landscape of product
development. Five robustness indicators were introduced \citep{juul2024multi} based on multi-objective design exploration to tackle two central challenges in engineering design: (1) defining design margins early in the design process, before detailed modeling is possible, and (2) evaluating interdependencies among those margins to prevent cascading effects. As a result, three distinct margin types were identified to help designers assess margin implications even before models are fully defined. A more holistic review of
margins throughout the product development process have been explored elsewhere \citep{brahma2024margins}. 

In comparison, although our approach is closely related to margin, our problem is different from those addressed in the current literature of margin. First, the current literature focus on the margin of design parameters while our approach aims to introduce margin into the domain and the co-domain of the functions. Furthermore, the concept of hierarchical refinement is relatively rare in the current literature. As recently pointed out, the concept of hierarchy in the margin literature is not well understood \citep{brahma2024margins}. 
While Jacobson and Ferguson touched on this concept \citep{jacobson2023hierarchical}, the proposed process was largely informal. In fact, the authors admitted at the end of the paper that better modeling approaches around decomposition are still needed. 
Finally, although this paper is not originally aimed to contribute to the margin community, we recognize that many concepts utilized in this paper indeed resemble margin (see Section~\ref{sec:discussion}). 
We believe the formal framework proposed by this paper can contribute to the vision of a design environment that will eventually connect margin to requirements \citep{ferguson2024managing}.

\vspace{11pt}
\noindent
Recall that this paper sets out to address two challenges: first to create a rigorous method for functional requirement decomposition and then to provide a verification framework for such decompositions. For Objective 1: There is little attention in the SBD literature to a general formal method for functional requirements decomposition in engineered systems. \citet{shallcross2020set} pointed out that ``there is limited SBD research contributing to requirements development,'' and noted that SBD methodologies applied to complex systems are mostly qualitative \citep{shallcross2021quantitative}. \citet{ghosh2014set} posited that SBD had not been formally defined, despite many authors studying its process inspired by the example of Toyota.

\section{Preliminaries}\label{sec:pre}
In general, functions describe the behavior between inputs and outputs of a system, subsystem, or component. Such a description can be an informal summary or statement, or a formal specification, all the way up to a literal mathematical {\em function} prescribing the desired input-output behavior . In the following, we take the latter perspective: the function of a system is a mathematical input-output transformation that describes the overall behavior of a system \citep{larson2012design, buede1997integrating}. 
\begin{equation}\label{eq:abstract}
 (x,u,c)\xrightarrow[]{f}y
\end{equation}

\begin{itemize}
    \item $f$ represents the transformation between $(x,u,c)$ and $y$.
    \item $x$ represents the input variables that the system takes from the environment or other systems.
    \item $y$ represents the output variables that the system gives in response to the inputs.
    \item $u$ represents the uncertain design parameters that are out of the designer's control, such as the weather for an airplane. We call them the uncontrollable parameters. 
    \item $c$ represents the uncertain design parameters that are under the designer's control, such as the weight of an airplane. We call them the controllable parameters. 
\end{itemize}
Note that $x, y, u$ and $c$ are all column vectors of $f$, as a function likely have multiple input variables, output variables, controllable parameters and uncontrollable parameters. 
\subsection{Notation}\label{sec:notation}
We introduce the notation used throughout this paper.

\begin{itemize}
   \item $f$ and $f_i$: $f$ is the higher level function to be decomposed into multiple sub-functions $f_i$ where $i=1,2,...,m$. 
    \item $a(i)$: As mentioned above, $x, y, u$ and $c$ are the four column vectors of $f$. Similarly, the four column vectors of sub-function $f_i$ are denoted as  $x(i), y(i), u(i)$ and $c(i)$. Abstractly, we denote these column vectors with a unified expression $a(i)$, where   
    $a(i)=(a_1(i),a_2(i),\dots,a_n(i))^\top$.
    \item $a_j(i)$, as the $j^{th}$ element of $a(i)$, represents one individual input/output variable or controllable/uncontrollable parameter of sub-function $f_i$ (where $j=1,2...,n$).

    \item $R_{a_j(i)}$ is the range of $a_j(i)$, i.e., $a_j(i)\in R_{a_j(i)}$.
    \item $\{a(i)\}$ is a set representation of $a(i)$, i.e., $\{a(i)\}=\{a_1(i),a_2(i),\dots,a_n(i)\}$.
    \item $R_{a(i)}=(R_{a_1(i)},R_{a_2(i)},\dots,R_{a_n(i)})^\top$ is a column vector of the ranges of $a(i)$.
    \item $\{R_{a(i)}\}$ is a set representation of $R_{a(i)}$, i.e., $\{R_{a(i)}\}=\{R_{a_1(i)},R_{a_2(i)},\dots,R_{a_n(i)}\}$.
\end{itemize}

We now define the following set operations, where $a$ and $b$ are two column vectors.

\begin{itemize}
    \item $\{a\}\sqcup \{b\}$ returns the union of the identifiers (instead of the values) of all the elements in $\{a\}$ and $\{b\}$. The identical variable between $\{a\}$ and $\{b\}$ is merged into one element. By ``identical'', we mean the elements that have the same physical meaning, the same data source, and hence the same value \emph{at all times}.
    \item $\{a\}\sqcap \{b\}$ returns the identifier of the identical variable(s) between $\{a\}$ and $\{b\}$.
    \item $\{a\}\sqsubseteq \{b\}$ returns \textit{true} if the identifiers (instead of the values) of all the variables in $\{a\}$ are included in $\{b\}$.
    \item $\{R_a\}\Cup \{R_b\}$ returns the set of the ranges of $\{a\}\sqcup \{b\}$. Because the identical elements between $\{a\}$ and $\{b\}$ are subject to the ranges in both $\{R_a\}$ and $\{R_b\}$, the resulting ranges for the identical elements are the intersections ($\cap$) of the respective ranges in $\{R_a\}$ and $\{R_b\}$.
    \item $a_j|\{R_b\}$ identifies the element in $\{b\}$ that is identical to $a_j$ and returns its range in $\{R_b\}$.
    \item $\mathcal{TR}(\{R_a\})$ transforms the set $\{R_a\}$ into the column vector $R_a$, i.e., $\mathcal{TR}(\{R_a\})=R_a$.
\end{itemize}

\subsection{Functional Requirements}\label{sec:fq}

Mathematically, a system function, as defined in Eq.~(\ref{eq:abstract}) must satisfy Eq.~(\ref{eq:requirements1}), where $(R^*_x,R^*_u,R^*_c, R^*_y)$ constitute the functional requirements of the system.
\begin{equation}\label{eq:requirements1}
    \forall x\in R^*_x, \forall u\in R^*_u, \exists c\in R^*_c: (x,u,c)\xrightarrow[]{f}y\in R^*_y
\end{equation}

\begin{itemize}
    \item $\forall R^*_x$ implies the system must be able to process all the values from the input set.
    \item $\forall R^*_u$ is because $u$ is out of the designer's control. The system must be able to process all possible $u$.
    \item $\exists R^*_c$ is because $c$ is under the designer's control. The designer only needs to find one $c$ within $R^*_c$ to satisfy the functional requirements.
    \item $R^*_y$ is the co-domain of $f$. All possible output values must be bounded within $R^*_y$.
\end{itemize}

However, not any random $(R^*_x,R^*_u,R^*_c, R^*_y)$ can be the requirements of $f$. $(R^*_x,R^*_u,R^*_c)$ and $R^*_y$ must satisfy Eq.~(\ref{eq:fq}), which implies that the functional requirements of a system are actually a constrained mapping between $(R^*_x,R^*_u,R^*_c)$ and $R^*_y$.
\begin{equation}\label{eq:fq}
    (R^*_x,R^*_c,R^*_u)\longrightarrow R^*_y \text{ s.t. Eq.~(\ref{eq:requirements1})}
\end{equation}

Similarly, the requirements of $f_i$ can be represented as:
\begin{equation}\label{eq:subfq}
    (R^*_{x(i)},R^*_{c(i)},R^*_{u(i)})\longrightarrow R^*_{y(i)} \text{ s.t. Eq.~(\ref{eq:requirements1})}
\end{equation}

Therefore, \textbf{the functional requirements decomposition problem} is to decompose the top-level requirements in Eq.~(\ref{eq:fq}) into a set of sub-requirements in Eq.~(\ref{eq:subfq}) at the lower level, repeat the same process independently at each level of abstraction, and eventually lead to a system that satisfies the top-level requirements in Eq.~(\ref{eq:fq}).

\section{Parallel development}\label{sec:pd}
We explain in this that refinement and composibility are the two constraints to ensure parallel development (i.e., Objective 2). 
\subsection{Refinement and Composability}\label{sec:property}

Refinement and composability play an important role in producing traceable requirements.

\subsubsection{Refinement}

\begin{figure}[!t]
\centerline{
\includegraphics[width=.6\textwidth]{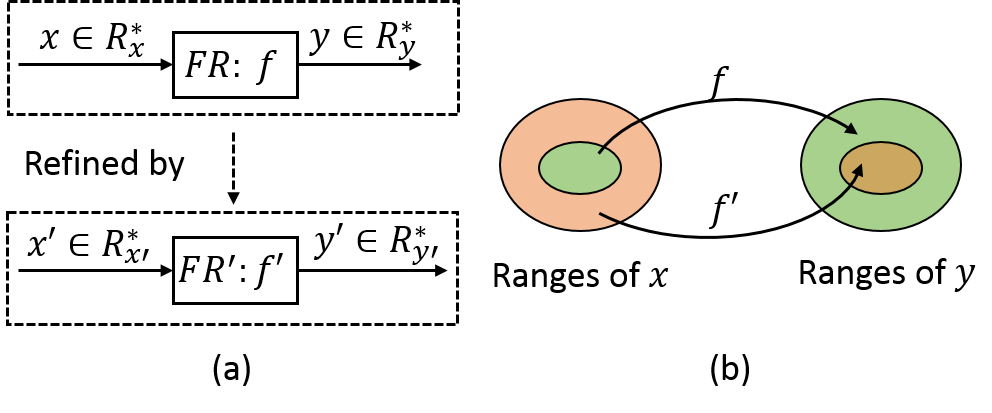}
}
\caption{Graphical illustration of \emph{refinement}. The green circles are the ranges defined in $FR$, and the brown circles are the ranges defined in $FR'$.}
\label{fig:refine}
\end{figure}

\textbf{Refinement} defines a relationship between two sets of functional requirements (FR). $FR'$ refining $FR$ implies the reduction of the uncertainties from $FR$ to $FR'$, which can always be reflected in the mapping between the input and output variables. As shown in Fig.~\ref{fig:refine}(a), $FR'$ refines $FR$ if the relationship in Eq.~(\ref{eq:refine}) is satisfied, where the meanings of $x_i| \{R^*_{x'}\}$ and $y_i|\{R^*_{y'}\}$ are defined in Section \ref{sec:notation}.

\begin{equation}\label{eq:refine}
\begin{cases}
     \{x\}\sqsubseteq \{x'\}\bigwedge \forall x_i\in \{x\}: R^*_{x_i}\subseteq x_i| \{R^*_{x'}\}\\
    \{y\}\sqsubseteq \{y'\}\bigwedge \forall y_i\in \{y\}: y_i|\{R^*_{y'}\}\subseteq R^*_{y_i}
\end{cases}
\end{equation}

Intuitively, Eq.~(\ref{eq:refine}) means (1) $FR'$ defines all the input/output variables of $FR$, and (2) $FR'$ allows the system to take a larger range of $x$ but will generate a smaller range of $y$ than $FR$.

Note that a system satisfying the functional requirements is a special case of \emph{refinement}. As implied by Eq.~(\ref{eq:requirements1}), the system can take all the values in $R^*_x$ and provide the output values that are bounded within $R^*_y$, which satisfies the relationship of \emph{refinement} in Eq.~(\ref{eq:refine}).

\subsubsection{Composability}

\begin{figure*}[!t]
\centerline{\includegraphics[width=\textwidth]{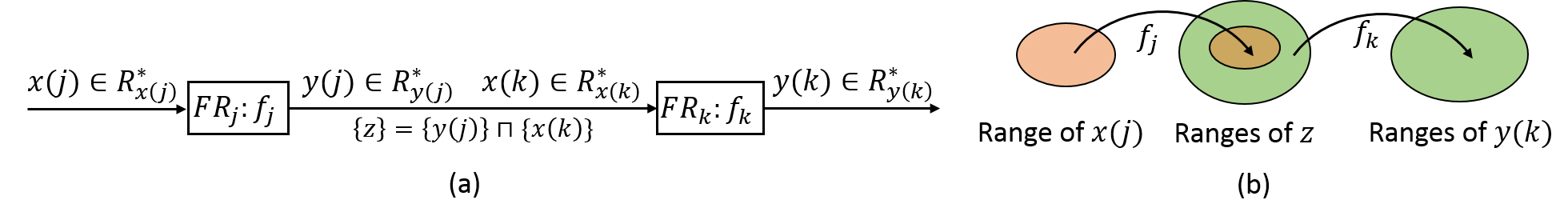}}
\caption{Graphical illustration of \emph{composability}. The green circles are the ranges defined in $FR_j$, and the brown circles are the ranges defined in $FR_k$.}
\label{fig:compose}
\end{figure*}

\textbf{Composability} defines the interface between functional requirements. Therefore, only the input and output variables are concerned. Two sets of functional requirements (denoted as $FR_j$ and $FR_k$ in Fig.~\ref{fig:compose}(a)) are composable if the relationship in Eq.~(\ref{eq:compose}) below is satisfied.
\begin{equation}\label{eq:compose}
    \begin{cases}
         \{z\}=\{y(j)\}\sqcap \{x(k)\}\neq \emptyset\\
         \forall z_i\in \{z\}: z_i|\{R^*_{y(j)}\}\subseteq z_i|\{R^*_{x(k)}\}
    \end{cases}
\end{equation}

Intuitively, \emph{composability} means two things: (1) the output of $FR_j$ and the input of $FR_k$ share at least one common variable (denoted as $\{z\}$), and (2) $FR_k$ can take in more values of $z$ than $FR_j$ can output (Fig.~\ref{fig:compose}(b)).

\subsection{Properties of Refinement and Composability}

Based on the definitions above, we have the following properties. %

\begin{property}\label{prop:p1}
If $FR'$ refines $FR$, and $FR''$ refines $FR'$, then $FR''$ refines $FR$.
\end{property}

\begin{property}\label{prop:p2}
If a set of functional requirements $\{FR_1, FR_2,\dots, FR_n\}$ are composable, then their refinements $\{FR'_1, FR'_2,\dots, FR'_n\}$ are also composable.
\end{property}

\begin{property}\label{prop:p3}
Consider a set of composable functional requirements $\{FR_1, FR_2,\dots, FR_n\}$. If $FR'_i$ refines $FR_i$ (i=1,2,\dots,n), then the composite of $\{FR'_1, FR'_2,\dots, FR'_n\}$ refines the composite of $\{FR_1, FR_2,\dots, FR_n\}$.
\end{property}

Because a system satisfying the functional requirements is a special case of \emph{refinement}, Property \ref{prop:p4}, \ref{prop:p5}, and \ref{prop:p6} can be derived similarly to Property \ref{prop:p1}, \ref{prop:p2}, and \ref{prop:p3}.

\begin{property}\label{prop:p4}
If $FR'$ refines $FR$, and a system $S$ satisfies $FR'$, then $S$ satisfies $FR$.
\end{property}

\begin{property}\label{prop:p5}
If a set of functional requirements $\{FR_1, FR_2,\dots, FR_n\}$ are composable, and a set of systems $\{S_1, S_2,\dots,S_n\}$ satisfy the functional requirements respectively, then $\{S_1, S_2,\dots,S_n\}$ are also composable.
\end{property}

\begin{property}\label{prop:p6}
Consider a set of composable functional requirements $\{FR_1, FR_2,\dots, FR_n\}$ and their composite $FR$. If $S_i$ satisfies $FR_i$ (i=1,2,\dots,n), then the composite of $\{S_1, S_2,\dots, S_n\}$ satisfies the composite of $\{FR_1, FR_2,\dots, FR_n\}$.
\end{property}

\subsection{Ensuring parallel development}\label{sec:independent}

\begin{figure}[!t]
\centerline{\includegraphics[width=.6\textwidth]{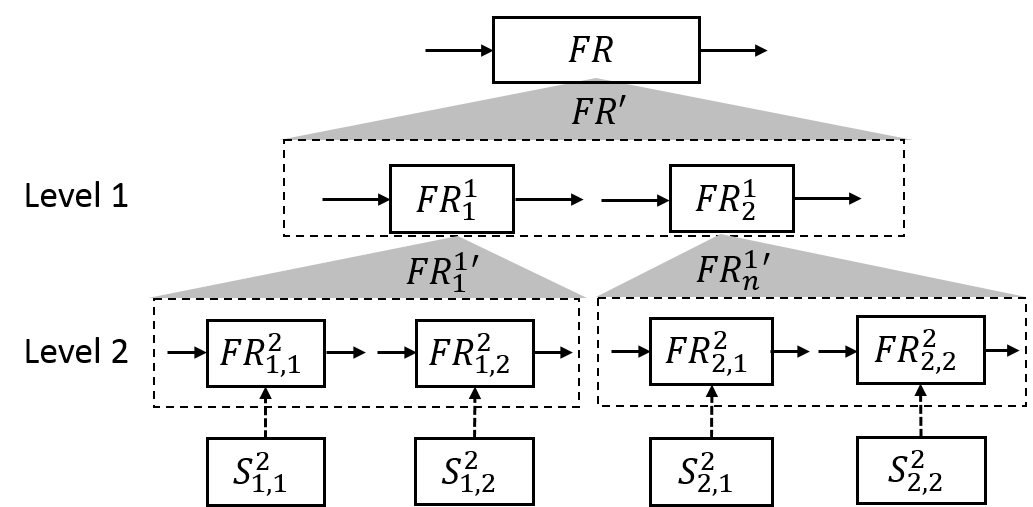}}
\caption{A two-level example system to describe the hierarchical functional requirements decomposition process.}
\label{fig:decomposition}
\end{figure}

In this subsection, we illustrate, with a simple system (Fig.~\ref{fig:decomposition}), the hierarchical functional requirements decomposition process and prove the conditions under which the decomposed requirements ensure the top-level functional requirements are satisfied.

First, the given top-level functional requirements $FR$ are decomposed into $\{FR_1^1, FR_2^1\}$ and assigns them to the Level-1 suppliers. The Level-1 suppliers then independently decompose $FR_1^1$ and $FR_2^1$ into $\{FR_{1,1}^2, FR_{1,2}^2\}$ and $\{FR_{2,1}^2, FR_{2,2}^2\}$, respectively, and assign these to Level-2 suppliers. The Level-2 suppliers implement the functional requirements independently into individual systems $S_{1,1}^2, S_{1,2}^2, S_{2,1}^2$, and $S_{2,2}^2$ that satisfy the respective functional requirements.

We now prove that if at each level of abstraction, (1) the sub-requirements are \emph{composable}, and (2) when composed together, the sub-requirements \emph{refine} the higher-level functional requirements, then $S_{1,1}^2, S_{1,2}^2, S_{2,1}^2$, and $S_{2,2}^2$ are composable and together satisfy $FR$.

\vspace{2mm}
\noindent\emph{Proof:} 

\begin{enumerate}
    \item By definition, $\{FR_1^1, FR_2^1\}$ are composable and refine $FR$; $\{FR_{1,1}^2, FR_{1,2}^2\}$ are composable and refine $FR_1^1$; $\{FR_{2,1}^2, FR_{2,2}^2\}$ are composable and refine $FR_2^1$; $S_{1,1}^2, S_{1,2}^2, S_{2,1}^2$, and $S_{2,2}^2$ satisfy $FR_{1,1}^2, FR_{1,2}^2, FR_{2,1}^2$, and $FR_{2,2}^2$, respectively.

    \item According to Property \ref{prop:p2}, $\{FR_{1,1}^2, FR_{1,2}^2, FR_{2,1}^2, FR_{2,2}^2\}$ are composable. Then, the composite of $\{FR_{1,1}^2, FR_{1,2}^2, FR_{2,1}^2, FR_{2,2}^2\}$ refines the composite of $\{FR_1^1, FR_2^1\}$ according to Property \ref{prop:p3}. Because $\{FR_1^1, FR_2^1\}$ refine $FR$, the composite of the form $\{FR_{1,1}^2, FR_{1,2}^2, FR_{2,1}^2, FR_{2,2}^2\}$ refines $FR$ according to Property \ref{prop:p1}.

    \item Since $\{FR_{1,1}^2, FR_{1,2}^2, FR_{2,1}^2, FR_{2,2}^2\}$ are composable, and $S_{1,1}^2, S_{1,2}^2, S_{2,1}^2$, and $S_{2,2}^2$ satisfy $FR_{1,1}^2, FR_{1,2}^2, FR_{2,1}^2$, and $FR_{2,2}^2$, respectively, then $\{S_{1,1}^2, S_{1,2}^2, S_{2,1}^2, S_{2,2}^2\}$ are composable according to Property \ref{prop:p5}. The composite of $\{S_{1,1}^2, S_{1,2}^2, S_{2,1}^2, S_{2,2}^2\}$ satisfies the composite of $\{FR_{1,1}^2, FR_{1,2}^2, FR_{2,1}^2, FR_{2,2}^2\}$ according to Property \ref{prop:p6}. We have established that the composite of $\{FR_{1,1}^2, FR_{1,2}^2, FR_{2,1}^2, FR_{2,2}^2\}$ refines $FR$. Then, according to Property \ref{prop:p4}, the composite of $\{S_{1,1}^2, S_{1,2}^2, S_{2,1}^2, S_{2,2}^2\}$ satisfies $FR$. In other words, the resulting system satisfies the top-level functional requirements. \qed
\end{enumerate}

Therefore, for design teams at different levels of abstraction to work in parallel and obtain a system that satisfies the top-level functional requirements, the sub-requirements at each level of abstraction must be \emph{composable} and \emph{refine} the higher-level functional requirements as a whole.

\section{Set-Based Requirements Decomposition}\label{sec:decompose}
In this section, we explain the set-based requirements decomposition process to tackle ambiguity and uncertainty in requirements decomposition (i.e., Objective 1).
\subsection{An overall description}
Given the higher-level functional requirements, the decomposition process involves the following steps:
\begin{enumerate}
    \item \emph{Define and analyze the functional architecture (Section \ref{sec:architecture}).} A set of functional architecture (comprised of a set of connected sub-functions) for the system-under-design are defined to refine the higher level functional requirements, based on which the design space and the performance space are derived. 
    \item \emph{Explore the sub-functions for initial feasible spaces (Section \ref{sec:sub-functions}).}  A set of initial feasible spaces of the sub-functions are characterized individually by exploring the different feasible implementations for the corresponding sub-functions.
    \item \emph{Narrow the initial feasible spaces(Section \ref{sec:dsexp}).} 
    The feasible design space and the feasible performance space of the system-under-design are narrowed gradually by taking ``intersections'' based on the initial feasible spaces of the sub-functions.
    \item \emph{Determine the sub-requirements (Section \ref{sec:reqdev}).} The functional requirements of each sub-functions are defined based on the narrowed feasible design space and performance space so that all the sub-requirements are \emph{composable}, and when composed together, they can \emph{refine} the higher-level functional requirements. 
\end{enumerate}
After that, the team for each sub-function will improve the level of detail by independently decomposing the sub-functions into the lower levels iteratively until a final solution can be chosen.

\subsection{Define and Analyze the Functional Architecture}\label{sec:architecture}
A functional architecture is composed of a set of connected sub-functions ($f_1, f_2, \dots, f_n$) and represents a set of potential design solutions for the higher-level requirements. In SBD, there can be multiple alternative functional architectures.  However, from the perspective of formal reasoning, the process to derive the sub-requirements for any functional architecture is the same. Therefore, we assume in this paper only one functional architecture is defined in this step. In addition, the identification of the functional architecture relies on the designers understanding of the desired system \citep{benveniste2015contracts}. The techniques to identify functional architecture are out of the scope of this paper but can be found elsewhere \citep{penzenstadler2011desyre, functiondecomp,komoto2011theory}. 

Given a functional architecture, we extract the following two pieces of information: (1) which variable/parameter is shared with which function/sub-functions, and (2) the constitution of the design space and the performance space. These will be used to narrow the feasible spaces later.

\textbf{First}, we aggregate $x(i), y(i), c(i)$, and $u(i)$ of the sub-functions into $\{x'\}, \{y'\}, \{c'\}$, and $\{u'\}$ without considering the functional architecture. As defined in Section \ref{sec:notation}, the $\sqcup$ operation merges the identical variables/parameters in multiple sub-functions into one variable/parameter in the new set.
\begin{equation}\label{eq:initial}
\begin{cases}
     \{x'\}=\{x(1)\}\sqcup \{x(2)\}\sqcup \dots \sqcup \{x(n)\}\\ 
     \{y'\}=\{y(1)\}\sqcup \{y(2)\}\sqcup \dots \sqcup \{y(n)\}\\
     \{c'\}=\{c(1)\}\sqcup \{c(2)\}\sqcup \dots \sqcup \{c(n)\}\\
     \{u'\}=\{u(1)\}\sqcup \{u(2)\}\sqcup \dots \sqcup \{u(n)\}
\end{cases}
\end{equation}

\textbf{Second}, for the system to implement the higher-level functional requirements, all the elements defined in the higher-level function $f$ (i.e., $\{x\}, \{y\}, \{c\}$, and $\{u\}$) must also be defined in the functional architecture, i.e., 
\begin{equation}\label{eq:include}
    \{x\}\sqsubseteq \{x'\}\bigwedge \{c\}\sqsubseteq \{c'\}\bigwedge \{u\}\sqsubseteq \{u'\} \bigwedge\{y\}\sqsubseteq \{y'\}
\end{equation}

\textbf{Third}, we examine where the variables and the parameters are defined. The elements of $\{c'\}$ and $\{u'\}$ may or may not be defined in $\{c\}$ and $\{u\}$. We define the elements of $\{c'\}$ and $\{u'\}$ that are not defined in $\{c\}$ and $\{u\}$ as $\{\widetilde{c}\}$ and $\{\widetilde{u}\}$, and thus have
\begin{equation}\label{eq:uc}
\begin{cases}
 \{c'\}=\{\widetilde{c}\}\sqcup \{c\}\\
 \{u'\}=\{\widetilde{u}\}\sqcup \{u\}
\end{cases}
\end{equation}

\begin{figure}[!t]
\centerline{\includegraphics[width=.4\textwidth]{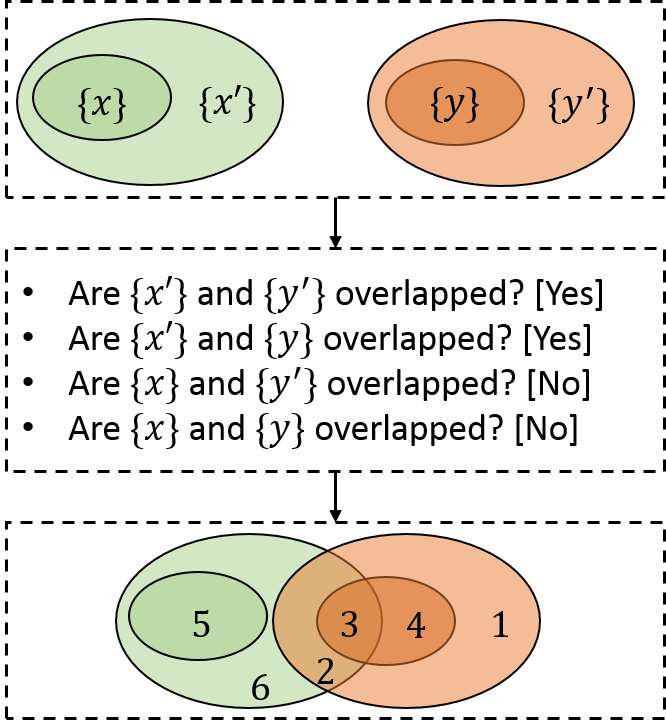}}
\caption{$\{x'\}$ and $\{y'\}$ can be divided into six groups to reason about the design space and the performance space. The ranges of each group will be calculated differently in Section \ref{sec:dsexp}.}
\label{fig:x'y'}
\end{figure}

For $\{x'\}$ and $\{y'\}$, we use Fig.~\ref{fig:x'y'} to reason about where the elements can be possibly defined. $\{x'\}$ and $\{y'\}$ are the two outer ovals that include $\{x\}$ and $\{y\}$. 

As shown in the middle of Fig.~\ref{fig:x'y'}, there are four possible different ways of where $\{x'\}$ and $\{y'\}$ can be defined. $\{x'\}$ and $\{y'\}$ overlap because the sub-functions are connected, i.e., one's input is another's output; $\{x'\}$ and $\{y\}$ overlap if there are feedback loops that feed some elements of $\{y\}$ into the sub-functions. $\{x\}$ does not overlap with $\{y'\}$ or $\{y\}$ because $\{x\}$ are the input variables of the higher-level functional requirements, hence can only come from the external environment rather than being controlled by the system itself. As a result, we identify six different areas (the bottom of Fig.~\ref{fig:x'y'}) depending on the overlapping between $\{x'\}, \{y'\}, \{x\}$, and $\{y\}$.

The variables in Areas 1, 2, 3, and 4 are therefore denoted as $\{y'^{(1)}\}, \{y'^{(2)}\}, \{y'^{(3)}\}$, and $\{y'^{(4)}\}$ in 
\begin{equation}\label{eq:4y'}
\begin{cases}
\{y'^{(1)}\}=\{y'\}\sqcap \neg \{y\}\sqcap \neg \{x'\}\\
\{y'^{(2)}\}=\{y'\}\sqcap \{x'\}\sqcap \neg \{y\}\\
\{y'^{(3)}\}=\{y\}\sqcap \{x'\}\\
\{y'^{(4)}\}=\{y\}\sqcap \neg \{x'\}
\end{cases}
\end{equation}
Area 5 is $\{x\}$. Area 6, denoted as $\{\widetilde{x}\}$, can then be represented as
\begin{equation}\label{eq:x''}
\{\widetilde{x}\} = \{x'\}\sqcap \neg \{y'\}\sqcap \neg \{x\}
\end{equation}

Therefore, we can classify all the elements of a functional architecture into the exclusive groups below based on where the elements are defined.
\begin{itemize}
    \item $\{x\}$ is defined in $\{x\}$ and $\{x'\}$; $\{\widetilde{x}\}$ is only defined within $\{x'\}$.
    \item $\{c\}$ is defined in $\{c\}$ and $\{c'\}$; $\{\widetilde{c}\}$ is only defined within $\{c'\}$.
    \item $\{u\}$ is defined in $\{u\}$ and $\{u'\}$; $\{\widetilde{u}\}$ is only defined within $\{u'\}$.
    \item $\{y'^{(1)}\}$ is only defined in $\{y'\}$; $\{y'^{(2)}\}$ is defined in $\{x'\}$ and $\{y'\}$; $\{y'^{(3)}\}$ is defined in $\{x'\}, \{y\}$, and $\{y'\}$; $\{y'^{(4)}\}$ is defined in $\{y\}$ and $\{y'\}$.
\end{itemize}

\textbf{Finally}, because the design space is the set of the independent variables/parameters of a system and the performance space is the set of the dependent variables \citep{wood1989computations,nahm2006novel}, the design space and the performance space of the therefore can be summarized in Eq.~(\ref{eq:dsps}).
\begin{equation}\label{eq:dsps}
    \begin{cases}
    \text{Design space: }\{x\}\sqcup\{\widetilde{x}\}\sqcup\{c\}\sqcup\{\widetilde{c}\}\sqcup \{u\}\sqcup\{\widetilde{u}\}\\
    \text{Performance space: }\{y'^{(1)}\}\sqcup \{y'^{(2)}\}\sqcup \{y'^{(3)}\}\sqcup \{y'^{(4)}\}
    \end{cases}
\end{equation}

\subsection{Explore the Initial Feasible Spaces}\label{sec:sub-functions}

The initial feasible spaces of each sub-function are characterized in 
\begin{equation}\label{eq:si}
\forall x(i)\in R_{x(i)}, \forall u(i)\in R_{u(i)}, \exists c(i)\in R_{c(i)}: f_i(x(i), u(i), c(i))\in R_{y(i)}
\end{equation}
after exploring the feasible implementations. The specific initial spaces can only be determined case by case.

Note that Eq.~(\ref{eq:si}) is different from Eq.~(\ref{eq:subfq}). The ranges (without $*$) in Eq.~(\ref{eq:si}) represent the feasible spaces of the sub-functions, while the ranges in Eq.~(\ref{eq:subfq}) represent the desired spaces (i.e., requirements).

\subsection{Narrow the Feasible Spaces}\label{sec:dsexp}

In this subsection, we derive the feasible design space and performance space of the system by narrowing from the initial feasible spaces.

\textbf{First}, we aggregate the ranges of the sub-functions $\{R_{x(i)}\}, \{R_{c(i)}\}, \{R_{u(i)}\}$, and $\{R_{y(i)}\}$ ($i=1,2,\dots,n$) into $\{R_{x'}\}, \{R_{c'}\}, \{R_{u'}\}$, and $\{R_{y'}\}$, where $\{x'\}, \{u'\}, \{c'\}$, and $\{y'\}$ are defined in Eq.~(\ref{eq:initial}). Refer to Section \ref{sec:notation} for the definition of $\Cup$.
\begin{equation}\label{eq:ranges}
\begin{cases}
       \{R_{x'}\}=\{R_{x(1)}\}\Cup \{R_{x(2)}\}\Cup \dots \Cup \{R_{x(n)}\}\\
       \{R_{y'}\}=\{R_{y(1)}\}\Cup \{R_{y(2)}\}\Cup \dots \Cup \{R_{y(n)}\}\\
    \{R_{c'}\}=\{R_{c(1)}\}\Cup \{R_{c(2)}\}\Cup \dots \Cup \{R_{c(n)}\}\\
    \{R_{u'}\}=\{R_{u(1)}\}\Cup \{R_{u(2)}\}\Cup \dots \Cup \{R_{u(n)}\}
\end{cases}
\end{equation}

\textbf{Second}, we calculate the ranges of $\{\widetilde{x}\}, \{x\}, \{\widetilde{c}\}, \{c\}, \{\widetilde{u}\}, \{u\}, \{y'^{(1)}\}, \{y'^{(2)}\}, \{y'^{(3)}\}$, and $\{y'^{(4)}\}$ based on where they are defined. For example, the elements in $\{x\}$ are defined in both $\{x\}$ and $\{x'\}$. Therefore, the range of each element in $\{x\}$ will be the intersection of the corresponding ranges in $R^*_{x}$ and $\{R_{x'}\}$, i.e., the first item in Eq.~(\ref{eq:xcut}). Refer to Section \ref{sec:notation} for the definition of the operation $x_i|\{R_{x'}\}$.

The ranges of $\{c\}, \{u\}, \{\widetilde{x}\}, \{\widetilde{c}\}, \{\widetilde{u}\}, \{y'^{(1)}\}, \{y'^{(2)}\}, \{y'^{(3)}\}$, and $\{y'^{(4)}\}$ can be calculated in the same way in the rest of Eq.~(\ref{eq:xcut}).
\begin{equation}\label{eq:xcut}
\begin{cases}
\{R_x\}=\{R_{x_i}|\forall x_i\in \{x\}:R_{x_i}=R^*_{x_i}\cap x_i|\{R_{x'}\} \}\\
 \{R_u\}=\{R_{u_i}|\forall u_i\in \{u\}:R_{u_i}=R^*_{u_i}\cap u_i|\{R_{u'}\} \}\\
  \{R_c\}=\{R_{c_i}|\forall c_i\in \{c\}:R_{c_i}=R^*_{c_i}\cap c_i|\{R_{c'}\} \}\\
   \{R_{\widetilde{x}}\}=\{R_{x_i}| \forall x_i\in \{\widetilde{x}\}: R_{x_i}=x_i|\{R_{x'}\}\}\\
      \{R_{\widetilde{c}}\}=\{R_{c_i}|\forall c_i\in \{\widetilde{c}\}:R_{c_i}= c_i|\{R_{c'}\} \}\\
     \{R_{\widetilde{u}}\}=\{R_{u_i}| \forall u_i\in \{\widetilde{u}\}: R_{u_i}=u_i|\{R_{u'}\}\}\\
      \{R_{y'^{(1)}}\}=\{R_{y_i}|\forall y_i\in \{y'^{(1)}\}: R_{y_i}=y_i|\{R_{y'}\}\}\\
        \{R_{y'^{(2)}}\} = \{R_{y_i}|\forall y_i\in \{y'^{(2)}\}:  R_{y_i} = y_i|\{R_{x'}\}\cap y_i|\{R_{y'}\} \}\\
    \{R_{y'^{(3)}}\} = \{R_{y_i}|\forall y_i\in \{y'^{(3)}\}:  R_{y_i} = y_i|\{R_{x'}\}\cap y_i|\{R_{y'}\}\cap y_i|\{R^*_y\}\}\\
     \{R_{y'^{(4)}}\} = \{R_{y_i}|\forall y_i\in \{y'^{(4)}\}:  R_{y_i} = y_i|\{R_{y'}\}\cap y_i|\{R^*_y\}\}       
\end{cases}
\end{equation}

In addition, each range in $\{R_x\}, \{R_u\}, \{R_c\}, \{R_{\widetilde{x}}\}, \{R_{\widetilde{c}}\}, \{R_{\widetilde{u}}\}, \{R_{y'^{(1)}}\}, \{R_{y'^{(2)}}\}, \{R_{y'^{(3)}}\}$, and $\{R_{y'^{(4)}}\}$ cannot be $\emptyset$ because $\emptyset$ implies internal conflicts between the initial ranges, unless the corresponding group of elements is not defined. Furthermore, for $\{R_x\}$ and $\{R_u\}$, the conditions are stricter. Recall $\forall$ is associated with $\{R^*_x\}$ and $\{R^*_u\}$ in Eq.~(\ref{eq:requirements1}). The system must operate under all the values defined in $\{R^*_x\}$ and $\{R^*_u\}$. For this reason, the ranges of $\{x\}$ and $\{u\}$ defined in $\{R_{x'}\}$ and $\{R_{u'}\}$ must include $\{R^*_x\}$ and $\{R^*_u\}$ respectively. As a result, we can update $\{R_x\}$ and $\{R_u\}$ in Eq.~(\ref{eq:xu}) based on Eq.~(\ref{eq:xcut}).
\begin{equation}\label{eq:xu}
\{R_x\}=\{R^*_x\} \bigwedge \{R_u\}=\{R^*_u\}
\end{equation}

Now, let $R^1_D=\{R_x\}\Cup \{R_{\widetilde{x}}\}\Cup \{R_{c}\}\Cup \{R_{\widetilde{c}}\}\Cup \{R_{u}\}\Cup \{R_{\widetilde{u}}\}$ and $R^1_P=\{R_{y'^{(1)}}\}\Cup \{R_{y'^{(2)}}\}\Cup \{R_{y'^{(3)}}\}\Cup \{R_{y'^{(4)}}\}$. The feasible design space and performance space can be easily represented as:
\begin{equation}\label{eq:fds1fps1}
\begin{aligned}
          FDS^1 = \mathcal{TR}(R^1_D) \bigwedge FPS^1 = \mathcal{TR}(R^1_P)
\end{aligned}
\end{equation}

\textbf{Third}, $FDS^1$ and $FPS^1$ are coupled through the sub-functions and can be further narrowed by exploiting the coupling. Let $f' = f_1 \wedge f_2 \wedge \dots \wedge f_n$. $f'$ can be characterized as:
\begin{equation}\label{eq:mapping}
(x, \widetilde{x}, c, \widetilde{c}, u, \widetilde{u}) \xrightarrow[]{f'} y'
\end{equation}
where $(x, \widetilde{x}, c, \widetilde{c}, u, \widetilde{u}) \in FDS^1$ and $y' \in FPS^1$.

Based on Eq.~(\ref{eq:mapping}), we can narrow $R_c$ and $R_{\widetilde{c}}$ to $R^{\downarrow}_c$ and $R^{\downarrow}_{\widetilde{c}}$ so that Eq.~(\ref{eq:inner}) is satisfied, which is a quantified constraint satisfaction problem \citep{qureshi2014set}.
\begin{equation}\label{eq:inner}
\text{For } \forall x \in R_{x}, \forall \widetilde{x} \in R_{\widetilde{x}}, \forall u \in R_{u}, \forall \widetilde{u} \in R_{\widetilde{u}}, \forall c \in R^{\downarrow}_{c}, 
\text{and } \forall \widetilde{c} \in R^{\downarrow}_{\widetilde{c}}: 
(x, \widetilde{x}, c, \widetilde{c}, u, \widetilde{u}) \xrightarrow[]{f'} y' \in FPS^1
\end{equation}
Let $R^2_D = \{R_x\} \Cup \{R_{\widetilde{x}}\} \Cup \{R_{u}\} \Cup \{R_{\widetilde{u}}\} \Cup \{R^{\downarrow}_{c}\} \Cup \{R^{\downarrow}_{\widetilde{c}}\}$, then the new feasible design space can be represented as $FDS^2 = \mathcal{TR}(R^2_D)$.

The reason that only $R_c$ and $R_{\widetilde{c}}$ are narrowed is because other ranges in Eq.~(\ref{eq:mapping}) are all associated with $\forall$, meaning the system must operate under all values in the ranges. Hence, ranges associated with $\forall$ cannot be narrowed. Furthermore, the reason that the ranges of the controllable parameters $R^{\downarrow}_c$ and $R^{\downarrow}_{\widetilde{c}}$ in Eq.~(\ref{eq:inner}) are associated with $\forall$ is that $R^{\downarrow}_c$ and $R^{\downarrow}_{\widetilde{c}}$ will be further narrowed independently by subsystems at the lower levels of abstraction until single values can be determined for the elements in $\{c\}$ and $\{\widetilde{c}\}$. ``$\forall$'' makes sure whatever values chosen for $c$ and $\widetilde{c}$ at the lower levels, Eq.~(\ref{eq:inner}) will always be satisfied.

After Eq.~(\ref{eq:inner}), because the feasible design space is narrowed from $FDS^1$ to $FDS^2$, $FPS^1$ can be narrowed into a new feasible performance space $FPS^2$ based on the new $FDS^2$, such that Eq.~(\ref{eq:task5}) holds:
\begin{equation}\label{eq:task5}
\forall (x, \widetilde{x}, c, \widetilde{c}, u, \widetilde{u})^\top \in FDS^2: (x, \widetilde{x}, c, \widetilde{c}, u, \widetilde{u}) \xrightarrow[]{f'} y' \in FPS^2
\end{equation}
Reachability analysis \citep{geretti2021arch} can be applied to calculate $FPS^2$ as an over-approximation so that all the points in $FDS^2$ will be mapped within $FPS^2$. Let the resulting reachable sets for $y'^{(h)}$ be $\{R_{y'^{(h)}}^\downarrow\}$ and $R_P^2 = \{R_{y'^{(1)}}^\downarrow\} \Cup \{R_{y'^{(2)}}^\downarrow\} \Cup \{R_{y'^{(3)}}^\downarrow\} \Cup \{R_{y'^{(4)}}^\downarrow\}$. The new feasible performance space can be represented as $FPS^2 = \mathcal{TR}(R^2_P)$.

Finally, the feasible design space and performance space of the system are narrowed to $FDS^2$ and $FPS^2$.

\subsection{Determine the Sub-Requirements}\label{sec:reqdev}

In this subsection, we determine the sub-requirements that are \emph{composable}, and when composed together, can \emph{refine} the higher-level functional requirements.

\paragraph{Uncertainty Expansion.} Ideally, the ranges in the sub-requirements should be as narrow as possible to reduce uncertainty to the greatest extent. The ranges in $FDS^2$ and $FPS^2$ are the narrowest ranges that can be calculated at the current level of abstraction. However, it is always possible that the initial characterizations of the uncertainties (i.e., $R_{u'}$ and $R_{c'}$) are under-estimated, which will inevitably lead to an expansion of $FPS^2$ at a later time.

\begin{figure}[!t]
\centerline{\includegraphics[width=.45\textwidth]{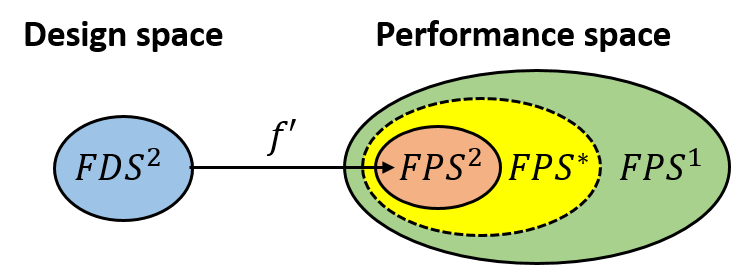}}
\caption{The boundaries of $FPS^*$.}\label{fig:fps}
\end{figure}

To accommodate the possible unexpected expansion, the desired ranges of $y'$ in the sub-requirements must also expand from $FPS^2$. We denote the desired ranges of $y'$ in the sub-requirements as $FPS^*$. As shown in Fig.~\ref{fig:fps}, to maximize the capacity of the sub-requirements to accommodate the possible expansion, $FPS^*$ must be expanded as close to $FPS^1$ as possible. Note that $FPS^*$ cannot expand beyond $FPS^1$ to avoid infeasibility.

However, $y'$ may also be the input variables of some sub-functions. Expanding the ranges of $y'$ will inevitably lead to the expansion of the input ranges of some sub-functions, which will in turn make the system more difficult or expensive to implement. Therefore, there is an inherent \textbf{tension} between accommodating excessive uncertainty and the level of difficulty in implementing the system. This tension dictates that the boundary of $FPS^*$ in Fig.~\ref{fig:fps} cannot be too close either to $FPS^1$ or to $FPS^2$.

\paragraph{Trade-off Study.} A trade-off needs to be made to decide the distance of $FPS^*$ from $FPS^1$ and $FPS^2$. This problem highly resembles the barrier function in optimization. Therefore, we construct an optimization problem based on the barrier function to conduct the trade-off study.

First, we construct the boundaries of the ranges in $FPS^*$. For $y_j(i)$ (i.e., the $j$th variable of $y(i)$), let the corresponding range defined in $FPS^*$ be $R^*_{y_j(i)}=[y_j^{l*}(i), y_j^{u*}(i)]$. As explained before, the ranges in $FPS^*$ must be bounded by the ranges in $FPS^1$ and $FPS^2$. Therefore, the range of $y_j(i)$ in $FPS^*$ is subject to Eq.~(\ref{eq:uplow1}).
\begin{equation}\label{eq:uplow1}
\begin{cases}
   y_j^{u2}(i) \leq y_j^{u*}(i) \leq y_j^{u1}(i)\\
   y_j^{l1}(i) \leq y_j^{l*}(i) \leq  y_j^{l2}(i)
\end{cases}
\end{equation}
\begin{itemize}
    \item $y_j^{u1}(i)$ and $y_j^{l1}(i)$ are the upper and lower bounds of the range of $y_j(i)$ in $FPS^1$.
    \item $y_j^{u2}(i)$ and $y_j^{l2}(i)$ are the upper and lower bounds of the range of $y_j(i)$ in $FPS^2$.
\end{itemize}

Second, we explain the barrier functions in Eq.~(\ref{eq:barrier}). $y_j(i)$ is the output of $f_i$. If $R^*_{y_j(i)}$ is too close to $FPS^2$, then $h_j(i)$ goes to infinity. If $y_j(i)$ is the input of $f_k$ and $R^*_{y_j(i)}$ is too close to $FPS^1$, then $h_j^k(i)$ goes to infinity; otherwise, $h_j^k(i)=0$.
\begin{equation}\label{eq:barrier}
\begin{cases}
    \begin{aligned}
   h_j(i) = -a_j(i)(&\ln|y_j^{u*}(i)-y_j^{u2}(i)| \\ + &\ln|y_j^{l*}(i)-y_j^{l2}(i)|)
\end{aligned}\\
    \begin{aligned}
  h_j^k(i) = -a_j^k(i)(&\ln|y_j^{u*}(i)-y_j^{u1}(i)| \\ + &\ln|y_j^{l*}(i)-y_j^{l1}(i)|)
\end{aligned}
\end{cases}
\end{equation}
The coefficients $a_j(i)$ and $a_j^k(i)$ represent the designers' preference for the ability to tolerate uncertainty expansion of $f_i$ and the level of difficulty in implementing $f_k$. Specific values of $a_j(i)$ and $a_j^k(i)$ can only be determined case by case.

\begin{equation}\label{eq:min}
    \text{min}\left(\sum_{i=1}^n \sum_{j=1}^{S(i)} h_j(i) + \sum_{i=1}^n \sum_{j=1}^{S(i)} \sum_{k=1}^n h_j^k(i)\right)
\end{equation}

Third, the objective function is Eq.~(\ref{eq:min}) where $S(i)$ is the number of the variables in $y(i)$. The optimization problem is to decide $y_j^{u*}(i)$ and $y_j^{l*}(i)$ to minimize Eq.~(\ref{eq:min}), subject to the constraints in Eq.~(\ref{eq:subfq}) and Eq.~(\ref{eq:uplow1}). As a result, $FPS^*$ can be assembled using the resulting $y_j^{u*}(i)$ and $y_j^{l*}(i)$ where $i=1,2,\dots,n$.

\paragraph{The Sub-Requirements.}
In general, the sub-requirements must satisfy the following three conditions:
\begin{enumerate}
    \item They must be within the initial feasible spaces of the sub-functions.
    \item They must include all the possible values that the elements of the system will encounter.
    \item They must be composable and refine the higher-level functional requirements as a whole.
\end{enumerate}
We intend to assign the ranges in $FDS^2$ and $FPS^*$ directly to the sub-functions as the sub-requirements. Now, we explain how the ranges satisfy all three conditions above.

First, $FDS^1$ and $FPS^1$ are both narrowed from the initial feasible spaces of the sub-functions (see Eq.~(\ref{eq:xcut})), and $FDS^2$ and $FPS^*$ are narrowed from $FDS^1$ and $FPS^1$. Therefore, the ranges in $FDS^2$ and $FPS^*$ satisfy the first condition.

Second, $FDS^2$ does not narrow the ranges of $\{x\}, \{\widetilde{x}\}, \{u\}$, and $\{\widetilde{u}\}$, elements that are associated with $\forall$, thus all the values of the independent variables that the system must accept are included within $FDS^2$. Moreover, we have established in Eq.~(\ref{eq:task5}) that all the possible values of $y'$ are included within $FPS^2$. Because $FPS^*$ includes $FPS^2$, all the possible values of $y'$ are also included within $FPS^*$. Therefore, the ranges in $FDS^2$ and $FPS^*$ satisfy the second condition.

Third, the ranges of all the elements are only defined once in $FDS^2$ and $FPS^*$. If they are assigned to the sub-functions, the identical element will only be assigned the same range. Therefore, the sub-requirements will be composable. In addition, the ranges of $\{x\}$ in $FDS^2$ are the same as the ranges in the higher-level functional requirements (see Eq.~(\ref{eq:xu})) and the ranges of $y$ are narrowed from the ranges in the higher-level functional requirements (see Eq.~(\ref{eq:xcut})). Therefore, the sub-requirements as a whole will also refine the higher-level functional requirement, and the third condition is satisfied.

In conclusion, $R^*_{x(i)}, R^*_{c(i)},$ and $R^*_{u(i)}$ can be directly obtained from $FDS^2$, and $R^*_{y(i)}$ can be directly obtained from $FPS^*$. Together, $R^*_{x(i)}, R^*_{c(i)}, R^*_{u(i)},$ and $R^*_{y(i)}$ are the sub-requirements of $f_i$ ($i=1,2,\dots,n$) decomposed from the $R^*_{x}, R^*_{c}, R^*_{u},$ and $R^*_{y}$ of $f$.

\section{Case Study}\label{sec:case}

\begin{figure}[htbp]
\centerline{\includegraphics[width=.4\textwidth]{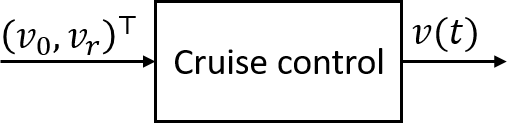}}
\caption{The top-level function.}
\label{fig:top}
\end{figure}

In this section, we design a cruise control system to demonstrate the four-step formal process proposed in Section \ref{sec:decompose}.
\paragraph{Problem description.} The task is to decompose functional requirements for a cruise control system. The top-level function can be represented as a mapping below, where the input is the initial speed $v_0$ and the reference speed $v_r$; the output is the real speed $v(t)$; no uncertainty parameters are defined. Fig.~\ref{fig:top} is a graphical representation of the mapping.
\begin{center}
    $(v_0,v_r)\xrightarrow[]{f} v(t)$
\end{center}

The requirements are summarized in Table \ref{tab:top-requirements}: for any initial speed $v_0\in [22.0,33.0]m/s$ and the reference speed $v_r\in [34.0,36.0]m/s$, the real speed $v(t)$ shall be bounded by $[20.0,40.0]m/s$ at all times and converge to $[33.0,37.0]m/s$ after 20 seconds.

\begin{table}[h]
    \centering
    \caption{Top-level functional requirements}
    \label{tab:top-requirements}
    \begin{tabular}{|c|c|p{1.4in}|}
\hline
                  & \textbf{Requirements}  & \textbf{Meaning} \\ \hline
\multirow{2}{*}{Input ($R_x^*$)} & $v_0 \in [22.0, 30.0]m/s$ & Given any initial speed in $[22.0, 30.0]m/s$ \\  \cline{2-3} 
                  & $v_r \in [34.0, 36.0] \ m/s$  & Given any desired speed in $[34.0, 36.0]m/s$ \\ \hline
\multirow{2}{*}{Output ($R_y^*$)} & $v(t)\in [20.0, 40.0]m/s, \ \forall t$  & The real speed is always bounded, i.e., stability \\ \cline{2-3} 
                  &$v(t) \in [33.0, 37.0] \ m/s, \ \forall t\in [20,\infty)$  & The real speed converges fast enough \\ \hline
             \makecell[c]{Controllable \\ parameters ($R_c^*$)}       & Not defined at current level & NA \\ \hline
            \makecell[c]{Uncontrollable \\ parameters ($R_u^*$)}& Not defined at current level & NA \\ \hline
\end{tabular}
\end{table}

\begin{figure}[htbp]
\centerline{\includegraphics[width=.6\textwidth]{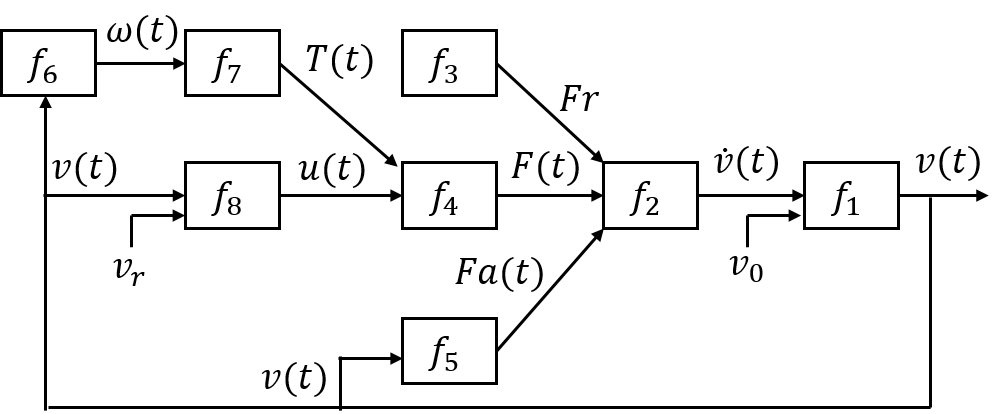}}
\caption{The functional architecture for the top-level functional requirements.}
\label{fig:designsolution}
\end{figure}

Intuitively, the task of decomposition is breaking down the top-level ``box'' in Fig.~\ref{fig:top} into more ``boxes'' at the lower level (to be explained in Fig.~\ref{fig:designsolution}).

\paragraph{Step 1: Define and analyze the functional architecture.} We adopt the functional architecture of a cruise control design from~\citep{fbs} (Fig.~\ref{fig:designsolution}) with each sub-function defined as follows:
\begin{equation*}
\begin{cases}
f_1: v(t) = \int_{0}^{t} \dot{v} \, dt + v_0 \\
f_2: \dot{v}(t) = \frac{F(t) - Fr - Fa(t)}{m} \\
f_3: Fr = mgC_r \\
f_4: F(t) = \alpha \cdot u(t) T(t) \\
f_5: Fa(t) = \frac{1}{2} \rho C_d A v(t)^2 \\
f_6: \omega(t) = \alpha \cdot v(t) \\
f_7: T(t) = T_m \left(1 - \beta \left(\frac{\omega(t)}{\omega_m} - 1\right)^2 \right) \\
f_8: u(t) = p(v_r - v(t)) + i \int_{0}^{t} (v_r - v(t)) \, dt
\end{cases}
\end{equation*}

Based on the architecture, we derive the design space and the performance of the cruise control system.
For example, according to the characterizations in Section \ref{sec:architecture}, $\{y'^{(3)}\}$ are the output variables that are defined in the output of the top-level function, the input of the sub-functions, and the output of the sub-functions. The only variable of such we can find is $v(t)$, and thus $\{y'^{(3)}\}=v(t)$. 
In addition, as defined in Fig.~\ref{fig:x'y'}, $\{y'^{(1)}\}$ and $\{y'^{(4)}\}$ are the output variables that are not the inputs to any sub-function.  In this example, no $\{y'^{(1)}\}$ or $\{y'^{(4)}\}$ is identified. 
Moreover, as depicted in Fig.~\ref{fig:x'y'}, $\{\widetilde{x}\}$ are the variables from the environment but are not defined in the top-level functional requirements. There is no such input variable in the functional architecture in Fig.~\ref{fig:designsolution}. Therefore, $\{\widetilde{x}\}$ is empty. Finally, because the uncertain parameters are not defined in the top-level requirements, $\{c\}$ and $\{u\}$ are empty. 

\begin{itemize}
    \item $\{x\} = \{v_r, v_0\}$.
    \item $\{\widetilde{c}\} = \{\omega_m\}$.
    \item $\{\widetilde{u}\} = \{m\}$.
    \item $\{y'^{(2)}\} = \{\dot{v}(t), Fr(t), F(t), Fa(t), T(t), u(t), \omega(t)\}$.
    \item $\{y'^{(3)}\} = v(t)$.
\end{itemize}
As a result, according to Eq.~(\ref{eq:dsps}), the design space of the cruise control system is, $$\{v_r, v_0, m, \omega_m\}$$ and the performance space of the cruise control system is, $$\{\dot{v}(t), Fr(t), F(t), Fa(t), T(t), u(t), \omega(t), v(t)\}.$$

\paragraph{Step 2: Explore the initial feasible spaces.}
The characterizations of the sub-functions depend on the feasible implementations of the respective sub-function, which can only be determined case by case. For this example, we characterize each sub-function of the functional architecture in Fig.~\ref{fig:designsolution} as below:

$\mathbf{f_1}$ is a mathematical definition of the speed, and hence will be allocated to no subsystem. $\dot{v}(t)$ has to be bounded to avoid violent acceleration/deceleration by the cruise control; $v(t)$ is bounded to constrain the fastest speed that the cruise control can be engaged due to safety concerns.  
\begin{itemize}
    \item $f_1$: $v(t)=\int_{0}^{t} \dot{v} dt+v_0$.
    \item $x(1)=(v_0,\dot{v})^\top$ and,\newline $R_{x(1)}=([0,40]\si[per-mode=symbol]{\meter\per\second},[-1.5,3]\si[per-mode=symbol]{\meter\per\second^2})^\top$. 
    \item $y(1)=v(t)$ and $R_{y(1)}=[0,50]\si[per-mode=symbol]{\meter\per\second}$.
\end{itemize}

$\mathbf{f_2}$ combines all the forces and yields the acceleration of the car as a whole, where $F(t)$ is the driving force, $Fa(t)$ is the aerodynamic drag, and $Fr$ is the tyre friction. The weight of the car $m$ is defined as an uncontrollable parameter.
\begin{itemize}
    \item $f_2$: $\dot{v}(t)=(F(t)-Fr-Fa(t))/m$.
    \item $x(2)=(F(t),Fr, Fa(t))^\top$ and,\newline
    $R_{x(2)}=([-250,3500]\si{N},[60,130]\si{N},[0,1000]\si{N})^\top$.
    \item $y(2)=\dot{v}(t)$ and $R_{y(2)}=[-2,4]\si[per-mode=symbol]{\meter\per\second^2}$;
    \item $u(2)=m$ and $R_{u(2)}=[990,1100]\si{kg}$.
\end{itemize}

$\mathbf{f_3}$ represent the rolling friction of the tyre. No input is defined for this function. $C_r$ is the rolling friction coefficient.
\begin{itemize}
    \item $f_3$: $Fr=mgC_r$.
    \item $y(3)=Fr$ and $R_{y(3)}=[70,120]\si{N}$.
    \item $u(3)=m$ and $R_{u(3)}=[990,1100]\si{kg}$.
\end{itemize}

$\mathbf{f_4}$ represents the ability of the drivetrain to convert the torque to the driving force, where $\alpha$ is the gear ratio, $u(t)$ is the control input from the cruise controller (not the uncontrollable uncertain parameters defined in this paper), $T(t)$ is the torque.
\begin{itemize}
    \item $f_4$: $F(t)=\alpha*u(t)T(t)$.
    \item $x(4)=(u(t),T(t))^\top$ and,\newline $R_{x(4)}=([-0.5,2],[0,250]\si{N\cdot m})^\top$.
    \item $y(4)=F(t)$ and $R_{y(4)}=[-1250,5000]\si{N}$.
\end{itemize}

$\mathbf{f_5}$ represents the aerodynamic drag, where $\rho$ is the air density, $C_d$ is
the shape-dependent aerodynamic drag coefficient and $A$ is the frontal area of the car.
\begin{itemize}
    \item $f_5$: $Fa(t)=\frac{1}{2}\rho C_dAv(t)^2$;
    \item $x(5)=v(t)$ and $R_{x(5)}=[0,60]\si[per-mode=symbol]{\meter\per\second}$.
    \item $y(5)=Fa(t)$ and $R_{y(5)}=[0,2000]\si {N}$.
\end{itemize}

$\mathbf{f_6}$ represents the ability of the drivetrain to convert from the car speed $v(t)$ to the engine speed $\omega(t)$.
\begin{itemize}
    \item $f_6$: $\omega(t)=\alpha*v(t)$;
    \item $x(6)=v(t)$ and $R_{x(6)}=[0,55]\si[per-mode=symbol]{\meter\per\second}$.
    \item $y(6)=\omega(t)$ and $R_{y(6)}=[0,560]\si[per-mode=symbol]{rad\per\second}$.
\end{itemize}

$\mathbf{f_7}$ represents the correlation between the torque, the engine speed $\omega(t)$, the maximum torque $T_m$ and the maximum engine speed $\omega_m$. $\omega_m$ is the controllable uncertain design parameter that starts with a rough estimation but will eventually be determined by the designers.
\begin{itemize}
    \item $f_7$: $T(t)=T_m(1-\beta(\omega(t)/\omega_m-1)^2)$.
    \item $x(7)=\omega(t)$ and  $R_{x(7)}=[0,450]\si[per-mode=symbol]{rad\per\second}$.
    \item $y(7)=T(t)$ and $R_{y(7)}=[0,250]$\si{NM}.
    \item $c(7)=\omega_m$ and $R_{c(7)}=[350,480]\si [per-mode=symbol]{rad\per\second}$.
\end{itemize}

$\mathbf{f_8}$ is the cruise controller, in this case based on PID control. $p$ and $i$ are the proportional and integral parameters for the PID controller. 
\begin{itemize}
    \item $f_8$: $u(t)=p(v_r-v(t))+i\int_{0}^{t}(v_r-v(t)) dt$.
    \item $x(8)=(v_r,v(t))^\top$ and,\newline $R_{x(8)}=([0,60]\si[per-mode=symbol]{\meter\per\second},[0,60]\si[per-mode=symbol]{\meter\per\second})^\top$.
    \item $y(8)=u(t)$ and $R_{y(8)}= [-0.5,4]$.
\end{itemize}

The coefficients mentioned above are $\alpha=10, g=9.8, C_r=0.01, C_d=0.32, \rho=1.3, A=2.4, \beta=0.4, T_m=200, p=0.1, i=0.5$.
Note that these constant coefficients can also be defined as uncertain parameters just like $m$ and $\omega_m$, which will make the computation more challenging. However, we intentionally limit the number of the uncertain parameters to to simplify the presentation.

\paragraph{Step 3: Narrow the feasible spaces.}
First, we compute $FDS^1$ and $FPS^1$ by taking intersections of the ranges of the elements defined in multiple places. For example, $m$ is defined in $f_2$ and $f_3$, hence $R_m$ is the intersection of the corresponding ranges defined in $f_2$ and $f_3$. $v_0$ is defined in both the top-level function and $f_1$, hence $R_{v_0}$ is the intersection of the corresponding ranges defined in $f_1$ and the top-level function.

As a result, the ranges in $FDS^1$ can be computed as follows:
\begin{center}
    $\begin{aligned}
    R_{v_r} &= [23.0, 30.0]\,\si{\meter\per\second}, \\
    R_{v_0} &= [34, 36]\,\si{\meter\per\second}, \\
    R_{\omega_m} &= [350, 480]\,\si{\radian\per\second}, \\
    R_m &= [990, 1100]\,\si{kg}
    \end{aligned}$
\end{center}
The ranges in $FPS^1$ can be computed as follows:
\begin{center}
    $\begin{aligned}
    R_{\dot{v}(t)} &= [-1.5, 3]\,\si{\meter\per\second^2}, \\
    R_{Fr(t)} &= [70, 120]\,\si{N}, \\
    R_{F(t)} &= [-250, 3500]\,\si{N}, \\
    R_{Fa(t)} &= [0, 1000]\,\si{N}, \\
    R_{T(t)} &= [0, 250]\,\si{Nm}, \\
    R_{u(t)} &= [-0.5, 2], \\
    R_{\omega(t)} &= [0, 450]\,\si{\radian\per\second}, \\
    R_{v(t)} &= [20, 40]\,\si{\meter\per\second}
    \end{aligned}$
\end{center}

Second, let $f' = f_1 \wedge f_2 \wedge \dots \wedge f_8$. $FDS^1$ and $FPS^1$ can be further narrowed into $FDS^2$ and $FPS^2$ using $f'$. To achieve $FDS^2$, the range of the controllable parameter $\omega_m$ must be narrowed into $R_{\omega_m}^{\downarrow}$ based on Eq.~(\ref{eq:inner}) such that,
\begin{center}
For $\forall v_r \in R_{v_r}, \forall v_0 \in R_{v_0}, \forall m \in R_{m}, \forall \omega_m \in R_{\omega_m}^{\downarrow}$:
\newline
   $(v_r, v_0, \omega_m, m) \xrightarrow[]{f'} y' \in FPS^1$
\end{center}  
where $y' = (v(t), \dot{v}(t), Fr(t), F(t), Fa(t), T(t), u(t), \omega(t))^\top$. In this example, we apply reachability analysis to help find $R_{\omega_m}^{\downarrow}$. We estimate an over-approximation of $R_{\omega_m}^{\downarrow}$, feed it into a reachability solver (i.e., CORA \citep{althoff2016cora} in this example), and adjust the bounds of the over-approximation until the reachable sets of all the output variables are bounded within $FPS^1$. 

As a result, we obtain the ranges in $FDS^2$ as follows:
\begin{center}
    $\begin{aligned}
    R_{v_r} &= [23.0, 30.0]\,\si{\meter\per\second}, \\
    R_{v_0} &= [34, 36]\,\si{\meter\per\second}, \\
    R_{\omega_m}^{\downarrow} &= [365, 450]\,\si{\radian\per\second}, \\
    R_m &= [990, 1100]\,\si{kg}
    \end{aligned}$
\end{center}

After that, $FPS^1$ can be narrowed into $FPS^2$ by calculating the reachable sets according to Eq.~(\ref{eq:task5}). We choose CORA as the tool because it can compute the over-approximation of the feasible performance space. The result shows the range of $v(t)$ converges gradually from $R_{v_0}$ to $R_{v_r}$, meaning the reference speed is achieved by the cruise control system. Eventually, we obtain the ranges of $FPS^2$:
\begin{center}
    $\begin{aligned}
    R_{\dot{v}(t)}^{\downarrow} &= [-0.63, 2.89]\,\si{\meter\per\second^2}, \\
    R_{Fr(t)}^{\downarrow} &= [88.20, 107.80]\,\si{N}, \\
    R_{F(t)}^{\downarrow} &= [-9.33, 2997.78]\,\si{N}, \\
    R_{Fa(t)}^{\downarrow} &= [240.88, 655.25]\,\si{N}, \\
    R_{T(t)}^{\downarrow} &= [179.94, 200.00]\,\si{Nm}, \\
    R_{u(t)}^{\downarrow} &= [0, 1.51], \\
    R_{\omega(t)}^{\downarrow} &= [219.66, 362.30]\,\si{\radian\per\second}, \\
    R_{v(t)}^{\downarrow} &= [21.97, 36.23]\,\si{\meter\per\second} \text{ for } [0, 100]\,\text{s}, \\
    R_{v(t)}^{\downarrow} &= [33.65, 36.19]\,\si{\meter\per\second} \text{ for } [20, 100]\,\text{s}
    \end{aligned}$
\end{center}
All the ranges in $FPS^2$ are subsets of their counterparts in $FPS^1$. In other words, $FPS^2$ is successfully narrowed from $FPS^1$.

\paragraph{Step 4: Determine the sub-requirements.}
We compute $R^*_{y(i)} = [y^{l*}(i), y^{u*}(i)]$ for $y(i)$ ($i=1,2,\dots,8$) by formulating the optimization problem in Eq.~(\ref{eq:min}) subject to the constraints in Eq.~(\ref{eq:subfq}) and Eq.~(\ref{eq:uplow1}). Note that there is no subscript $j$ associated with the output variable $y(i)$, because each $f_i$ in this example only has one variable at its output side.

First, the constraints described in Eq.~(\ref{eq:uplow1}) can be directly obtained from $FDS^1, FPS^1, FDS^2$, and $FPS^2$ calculated in the previous step.

Second, Eq.~(\ref{eq:subfq}) needs to be satisfied for each individual sub-function, which yields the constraints in Eq.~(\ref{eq:4}). Note that $f_1$ and $f_8$ are not included in the constraints. $f_1$ and $f_8$ alone are not enough to decide the ranges of the output variables because the operation $\int$ makes the ranges of the output variables correlated with the time $t$. Their ranges can only be determined after the rest of the system is determined. In this example, $f_1$ and $f_8$ belong to the PID controller. These modules are designed after other parts of the car (e.g., engine and tires) are defined, and hence are not part of the constraints in Eq.~(\ref{eq:4}).

\begin{equation}\label{eq:4}
\begin{cases}
y^{l*}(6) \leq f_6(y^{l*}(1)) \\ 
y^{u*}(6) \geq f_6(y^{u*}(1)) \\
y^{l*}(7) \leq f_7(y^{l*}(6), \overline{\omega_m}) \\ 
y^{u*}(7) \geq f_7(y^{u*}(6), \underline{\omega_m}) \\
y^{l*}(5) \leq f_5(y^{l*}(1)) \\
y^{u*}(5) \geq f_5(y^{u*}(1)) \\
y^{l*}(4) \leq f_4(y^{l*}(7), y^{l*}(8)) \\
y^{u*}(4) \geq f_4(y^{u*}(7), y^{u*}(8)) \\
y^{l*}(2) \leq f_2(y^{l*}(4), y^{u*}(3), y^{u*}(5), \overline{m}) \\
y^{u*}(2) \geq f_2(y^{u*}(4), y^{l*}(3), y^{l*}(5), \underline{m})
\end{cases}
\end{equation}

Third, based on Eq.~(\ref{eq:barrier}) and Eq.~(\ref{eq:min}), we define the objective function below:
\begin{equation}\label{eq:min1}
       \text{min} \left( \sum_{i=1}^8 h(i) + \sum_{i=1}^8 \sum_{k=1}^8 h^k(i) \right)
\end{equation}
where the coefficients are defined as $a(1)=0.9, a(2)=0.6, a(3)=0.5, a(4)=0.3, a(5)=0.4, a(6)=0.5, a(7)=0.1, a(8)=0.9, a^5(1)=0.5, a^6(1)=0.1, a^8(1)=0.8, a^1(2)=0.6, a^2(3)=0.4, a^2(4)=0.8, a^2(5)=0.7, a^7(6)=0.6, a^4(7)=0.9, a^4(8)=0.9$.

Finally, the optimization problem is to decide $y^{u*}(i)$ and $y^{l*}(i)$ ($i=1,2,\dots,8$) by minimizing Eq.~(\ref{eq:min1}). As a result, $y^{u*}(i)$ and $y^{l*}(i)$ ($i=1,2,\dots,8$) are calculated. Together with $FDS^2$ computed in the previous step, the ranges of each sub-function can be summarized in Table \ref{tab:results}. Accordingly, the sub-requirements of $f_i$ are the mapping between $(R^*_{x(i)}, R^*_{c(i)}, R^*_{u(i)})$ and $R^*_{y(i)}$ in Table \ref{tab:results}.

\begin{table}[htbp]
\centering
\caption{The requirements of the sub-functions.}
\label{tab:results}
\begin{tabular}{|l|l|l|}
\hline
 $f$ & $R_{x(i)}^*,R_{c(i)}^*,R_{u(i)}^*$ & $R_{y(i)}^*$  \\ \hline
 $f_1$  & \shortstack[l]{$v_0\in[34, 36]\si[per-mode=symbol]{\meter\per\second}$\\$\dot{v}(t)\in[-1.1, 3]\si[per-mode=symbol]{\meter\per\second^2}$} & $v(t)\in [21.8,38.4]\si[per-mode=symbol]{\meter\per\second}$ \\ \hline
 $f_2$  & \shortstack[l]{$Fr(t)\in[88.2, 107.8]\si{N}$\\
$F(t)\in[-159.1, 3024.0]\si {N}$\\
$Fa(t)\in[237.6, 827.6]\si{N}$\\$ m\in[990,1100]\si{kg}$} & $\dot{v}(t)\in[-1.1, 3]\si[per-mode=symbol]{\meter\per\second^2}$ \\ \hline
 $f_3$  & $m\in[990,1100]\si{kg}$ & $Fr(t)\in[88.2, 107.8]\si{N}$  \\ \hline
 $f_4$  &  \shortstack[l]{$T(t)\in[150, 200.1]\si{Nm}$\\$u(t)\in[-0.1, 1.511]$ }&$F(t)\in[-159.1, 3024.0]\si{N}$  \\ \hline
 $f_5$  & $v(t)\in [21.8,38.4]\si[per-mode=symbol]{\meter\per\second}$ & $Fa(t)\in[237.6, 827.6]\si{N}$ \\ \hline
 $f_6$  &$v(t)\in [21.8,38.4]\si[per-mode=symbol]{\meter\per\second}$  & $\omega(t)\in[109.8, 406.1]\si [per-mode=symbol]{\radian\per\second}$ \\ \hline
 $f_7$  &\shortstack[l]{$\omega(t)\in[109.8, 406.1]\si [per-mode=symbol]{\radian\per\second}$\\ $\omega_m\in[365,450]\si [per-mode=symbol]{\radian\per\second}$}  & $T(t)\in[150, 200.1]\si{Nm}$ \\ \hline
 $f_8$  & \shortstack[l]{$v_r\in[23.0,30.0]\si[per-mode=symbol]{\meter\per\second}$\\$v(t)\in [21.8,38.4]\si[per-mode=symbol]{\meter\per\second}$ } & $u(t)\in[-0.1, 1.511]$ \\ \hline
\end{tabular}
\end{table}

\begin{figure}[htbp]
\centerline{\includegraphics[width=.7\textwidth]{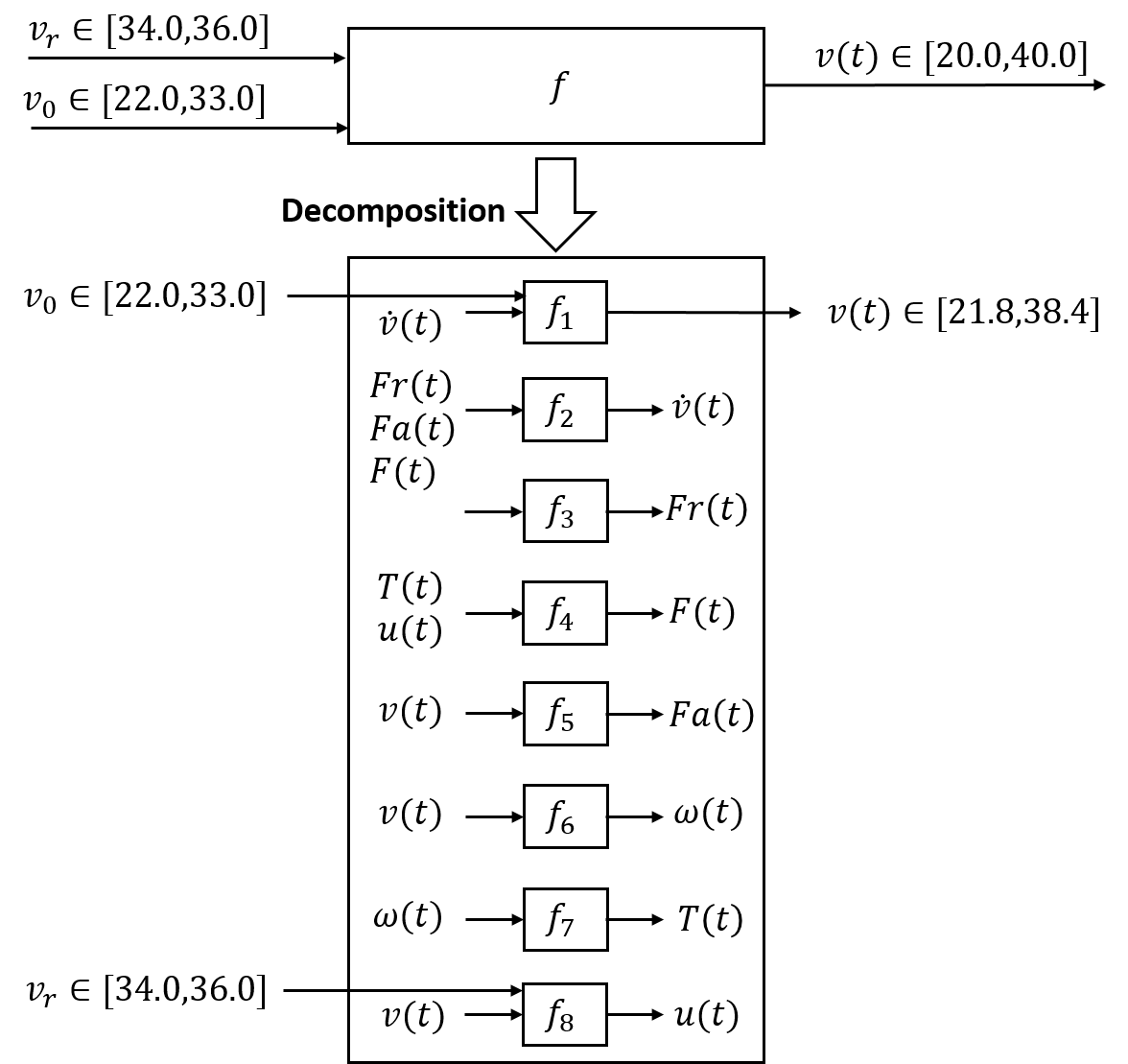}}
\caption{A graphical representation of the results.}
\label{fig:result}
\end{figure}

Fig.~\ref{fig:result} is a graphical representation of the results. The top-level function $f$ is decomposed into eight sub-functions $f_1,f_2,...,f_8$. As a whole, the eight functions have the same input variables and output variables as the top-level function. Other input variables of the sub-functions can be found at the outputs of the sub-functions, and vice versa. As we can see in Table~\ref{tab:results}, the variables that appear at both the input side and the output side have the same ranges, meaning the sub-functions are \emph{composable}. 
Furthermore, the sub-functions as a whole \emph{refine} $f$ because the input sets of the former include no less elements than the latter and the output set includes no more elements than the former. Based on our findings of parallel development in Section~\ref{sec:pd}, these sub-functions can be implemented independently. The same process can be repeated to further decompose the sub-functions at even lower levels, which eventually will lead to a system that satisfies the top-level functional requirements.

\section{Conclusion}
\subsection{Summary}\label{sec:summary}
Functional requirements decomposition at the early stage is challenging due to the uncertainty and ambiguity in design and the loose inter-team coordination especially in large projects for complex systems (e.g., commercial airlines).
SBD has demonstrated excellent potential in addressing these challenges but with limited formal guidance. To tackle uncertainty and ambiguity, one of the most effective approaches is to formalize. SBD's claim for team autonomy lacks formal proof as well. For this reason, this paper formalizes the functional requirements decomposition process based on SBD principles. We first proved parallel development: as long as the sub-requirements are composable and refine the higher-level functional requirements, the design teams at multiple levels can decompose their functional requirements independently, and the final system will still satisfy the top-level functional requirements. After that, we proposed a four-step formal process that supports parallel development to decompose the functional requirements based on SBD principles. Finally, a case study on designing a cruise control system was conducted to demonstrate the effectiveness of the proposed approach.

\subsection{Discussion}\label{sec:discussion}
\paragraph{Toyota's process.} According to Specking et al., much of the SBD literature shares similar characteristics with Toyota's process \citep{specking2018literature}.   
We compare our process with Toyota's approach below. In Toyota's approach  \citep{ward1995second}:

\begin{enumerate}
    \item The team defines a set of solutions at the system level.
    \item It defines sets of possible solutions for various subsystems.
    \item It explores these possible subsystems in parallel to characterize a possible set of solutions.
    \item It gradually narrows the set of solutions, converging slowly toward a single solution. In particular, the team determines the appropriate specifications to impose on the subsystems.
    \item Once the team establishes the single solution for any part of the design, it does not change it unless absolutely necessary.
\end{enumerate}
The first step above is the same as Step 1 in our process, where the ``solution'' in our case is the functional architecture. The second and third steps above correspond to Step 2 of our process, which essentially involves defining the feasible implementation of the subsystems individually. The fourth step above corresponds to Steps 3 and 4 in our process. However, our formal process provides more details without ambiguity. With the underlying formalism defined in the process, we can explain how to narrow the sets, determine the specifications on the subsystems, and gradually converge to a single solution using hierarchy. The fifth step above is related to the uncertainty expansion addressed in our process. Specifically, Toyota applies ``rigorous effort to avoid changes that \emph{expand}, rather than contract, the space of possible designs'' \citep{ward1995second}. We believe expansion is sometimes unavoidable due to epistemic uncertainties. Step 4 of our process for ``uncertainty expansion'' is motivated for this reason and designed to address the possible impacts of unexpected uncertainty expansions.

\paragraph{Implication to design.} Requirements decomposition is not a stand-alone activity in design. It is inherently multidisciplinary, integrating principles from systems engineering, software development, human factors, and organizational behavior. 
Our approach makes positive implications to the overall design by addressing the following four concerns while the design is progressing to one level lower: (1) the sub-requirements must not be technically too demanding, making them infeasible or too expensive to implement, (2) the sub-requirements must not be too limited, failing to support the intended system-level function and performance, (3) the sub-requirements must not be too relaxed, wasting too much capability of the subsystems, and (4) the sub-requirements should not be too tight, leaving little room for uncertainty. Traditionally, addressing these concerns requires intense communication efforts and conflict resolution between various stakeholders. Our approach solves them with a formal framework, increasing communication efficiency, directing conflict resolution, and ensuring successful integration with mathematical rigor. However, it is worth noting that formalism is by no means a silver bullet to solve all the challenges in requirements decomposition. 
Our approach should be blended with other activities (e.g., requirements negotiation and change management) and methods (e.g., optimization and reachability analysis) to eventually achieve a more effective and efficient requirements decomposition process.

\paragraph{Scope.} Our approach enhances requirements decomposition, but does not solve all the challenges in the requirements decomposition. First, requirements decomposition entails much more equally important topics than presented in this paper, such as requirements elicitation, architecture definition and requirements validation \& verification. This paper sets itself apart by focusing on addressing the four concerns discussed above. In the future, we will seek to understand how to integrate our approach with other equally important topics to further advance the research of requirements decomposition. Second, the proposed four-step process is a framework that focuses on articulating \emph{what} needs to be solved rather than \emph{how} to solve them. Multiple computational problems (e.g., optimization) are identified and formally defined without specifying the methods and tools to solve them and prescribing how well they need to be solved. Such ambiguity is left on purpose because (1) these computational problems on their own are still very active research topics, and (2) we intend to leave the most autonomy to the design teams to decide how to solve these problems with what tools. For this reason, we intentionally downplay the computational analysis (e.g., complexity analysis) in the paper as they are not directly relevant to addressing the four concerns discussed above and depend on the specific methods and tools selected by the engineers. We \emph{do not} claim a contribution to the computational aspect of the proposed process. 
Finally, formalism means precision but does not remove  engineer's judgment out of the equation.
Formalism in our framework instills precision in the decomposition process and reduces communication efforts by eliminating conflicts with the formally defined procedural requirements in place. 
But engineer's judgment are still required as inputs. For example, the coefficients of the barrier function in Eq.~(\ref{eq:barrier}) reflect engineer's tolerance for uncertainty expansion and their evaluation of the level of difficulty in implementing the sub-requirements. Determining these coefficients is highly subjective, same as many other optimization problems that involve human subjective inputs. We do not study how these coefficients are determined in this paper, but will explore the underlying decision-making mechanism in the future. 

\paragraph{Managing complexity.} A complex system usually have hundreds, if not thousands, of functions. Such a level of complexity is a main challenge for designing modern systems. As a solution, abstraction is one of the most effective tools to manage complexity \citep{hill2008applying,bencomo2024abstraction}. The activity of decomposition on one hand is refining the higher level abstraction, and on the other hand is making abstractions for the lower-level system. The reason for applying abstraction is because there is a limit to human's cognitive manageability. It is recommended by multiple systems engineering handbook \citep{incose2023incose,kapurch2010nasa} that decomposition should maintain a reasonable number of functions per level. This number is generally aligned with the famous Miller's law \citep{miller1956magical}: 5--9 functions per level, which is also the choice of the cruise control case study in this paper. Furthermore, the proposed process is agnostic to the number of sub-functions abstracted per level. The sub-functions can be further decomposed into multiple levels of smaller components. In fact, the deep hierarchical abstraction of the complex system makes the SBD-based method a particularly suitable approach to requirements decomposition.

\paragraph{Margin.} As mentioned before, many concepts incorporated in this paper are closely related to margin. First, in Eq.~(\ref{eq:requirements1}), $u$ denotes the uncertain parameters that cannot be controlled with $R_u^*$ as the a quantification of the uncertainty. While the uncertainty might be reduced due to for example a better understanding of the system and the environment, $\forall u\in R_u^*$ instill a margin (more precisely, buffer) into the system design at the current level that no matter whether and how the uncertainty is reduced, the uncertainty can always be absorbed by the design. Second, the $\in$ in Eq.~(\ref{eq:requirements1}) ensures all the output values generated from the inputs and parameters can be successfully processed by the system, creating a margin (more precisely, excess) for the system output. Third, $c$ and $\widetilde{c}$ of  Eq.~(\ref{eq:inner}) are controllable parameters, which are associated with $\exists$ in the original system  formulation in Eq.~(\ref{eq:requirements1}). But the narrowed bounds  $R^{\downarrow}_c$ and $R^{\downarrow}_{\widetilde{c}}$ in Eq.~(\ref{eq:inner}) are associated with $\forall$. This is because $c$ and $\widetilde{c}$ will be further narrowed at the lower level of abstraction. $\forall$ provides a margin (more precisely, buffer) for the lower levels so that no matter how $c$ and $\widetilde{c}$ are decided at a later time, it will never cause a violation at the higher level. Fourth, $FPS^2$ is computed in Eq.~(\ref{eq:task5}) as an over-approximation of $FDS^2$ due to technique difficulties in calculating the precise space of $y'$. Such an over-approximation formulates a margin (more precisely, excess) in $FPS^2$ w.r.t. $FDS^2$. Fifth and finally, uncertainty expansion in Section \ref{sec:reqdev} expands the performance space from $FPS^2$ to $FPS^*$ introducing a margin (more precisely, excess) for a potential growth of uncertainty.

\paragraph{Future work.} Our approach is only applicable to the systems that can be abstracted as input-output transformations (e.g., dynamic systems), while general SBD does not assume any formalism of the system.  This is a limitation of the paper. The applicability of other formalisms within the framework should be investigated in the future.